\documentclass[twocolumn,preprintnumbers,amsmath,amssymb,superscriptaddress,nofootinbib,longbibliography]{revtex4-1}

\usepackage{graphicx}
\usepackage{bm}
\usepackage{dsfont}
\usepackage[usenames,dvipsnames]{xcolor}
\usepackage{pstricks}
\usepackage[tight]{subfigure}
\usepackage{verbatim}
\usepackage{units}
\usepackage{multirow}
\usepackage{enumitem}
\usepackage{mathrsfs}
\usepackage{leftidx}
\usepackage{xspace}
\usepackage{braket}
\usepackage{bbm}
\usepackage{amsfonts,amssymb,amsmath}

\usepackage{psfrag}
\usepackage[abs]{overpic}
\usepackage{natbib}
\usepackage{float}
\usepackage[a4paper]{hyperref}
\hypersetup{colorlinks=true,linktoc=all,linkcolor=blue,breaklinks=true,citecolor=blue,urlcolor=blue}

\AtBeginDocument{%
    \newwrite\bibnotes
    \def\bibnotesext{Notes.bib}
    \immediate\openout\bibnotes=\jobname\bibnotesext
    \immediate\write\bibnotes{@CONTROL{REVTEX41Control}}
    \immediate\write\bibnotes{@CONTROL{%
    apsrev41Control,author="08",editor="1",pages="1",title="0",year="1"}}
     \if@filesw
     \immediate\write\@auxout{\string\citation{apsrev41Control}}%
    \fi
}%

\begin{document}

\title{Transport fluctuation relations in interacting quantum pumps}
\author{Roman-Pascal~Riwar}

\affiliation{Peter Gr\"unberg Institute: Theoretical Nanoelectronics,
Forschungszentrum J\"ulich, D-52425 J\"ulich, Germany}

\author{Janine~Splettstoesser}

\affiliation{Department of Microtechnology and Nanoscience, Chalmers University of Technology, S-41296 G\"{o}teborg, Sweden}

\date{\today}

\begin{abstract}

The understanding of out-of-equilibrium fluctuation relations in small open quantum systems has been a focal point of research in recent years. In particular, for systems with adiabatic time-dependent driving, it was shown that the fluctuation relations known from stationary systems do no longer apply due the geometric nature of the pumping current response. However, the precise physical interpretation of the corrected pumping fluctuation relations as well as the role of many-body interactions remained unexplored. Here, we study quantum systems with many-body interactions subject to slow time-dependent driving, and show that fluctuation relations of the charge current can in general not be formulated without taking into account the total energy current put into the system through the pumping process. Moreover, we show that this correction due to the input energy is nonzero only when Coulomb-interactions are present. Thus, fluctuation response relations offer an until now unrevealed opportunity to probe many-body correlations in quantum systems. We demonstrate our general findings at the concrete example of a single-level quantum dot model, and propose a scheme to measure the interaction-induced discrepancies from the stationary case.

\end{abstract}

\maketitle

\section{Introduction}\label{sec_intro}

Fluctuation relations of quantum observables~\cite{Callen1951Jul}, and in particular their generalization to nonequilibrium driven quantum systems~\cite{Esposito2006Apr,Andrieux2007Feb,Esposito2007Sep,Foerster2008Sep,Esposito2009Dec,Utsumi2010Mar,Ren2010Apr,Campisi2011Jul,Saira2012Oct,Lippiello2014Apr,Ansari2015Mar,Borrelli2015Jan,Cuetara2015May,Altaner2016Oct,Benito2016Nov}, are of fundamental importance to understand the second law of thermodynamics on a mesoscopic scale. The fluctuation relations themselves are formulated in terms of symmetries of the cumulant generating function. From these symmetries, a hierarchy of transport relations 
can be derived~\cite{Andrieux2007Feb}. They relate different individual cumulants of charge and energy currents of different orders in the response to a chemical potential or temperature gradient, which is why they are now commonly referred to as fluctuation-\textit{response} relations (FRR)~\cite{Altaner2016Oct}. 
From an application point of view, FRR can lead to powerful metrologic tools: for instance, the equilibrium fluctuation-dissipation theorem~\cite{Callen1951Jul,Kubo1966Jan} represents the centerpiece of Johnson thermometry (see Ref.~\cite{White1996Aug} and references therein). From a fundamental perspective, the symmetries leading to the fluctuation relations might potentially become important for a better understanding of nonequilibrium topological phase transitions~\cite{Ren2013May,Riwar2019Dec}.

In recent years, fluctuation relations for time-dependently driven systems have come to the focus of attention~\cite{Moskalets2009Aug,Ren2010Apr,Bulnes2014May,Croy2016Apr,Watanabe2017Aug,Hino2020Jul}. This is timely, since experiments have advanced to the point that not only average charge currents, but also their fluctuations - the noise - could be measured accurately~\cite{Flindt2009Jun,Parmentier2012Apr,Bocquillon2012May,Bocquillon2013Mar,Gabelli2013Feb,Dubois2013Oct,Jullien2014Oct}. Even counting of charge on quantum-dot pumps has been realized~\cite{Fricke2013Mar,Giblin2016Jan}. A main realization of this effort was, that fluctuation relations known from stationary systems do in general not simply extend to driven systems, due to the geometric nature of the system's response to the driving~\cite{Ren2010Apr}.

However, the previous theoretical works on fluctuation relations for quantum pumps~\cite{Ren2010Apr,Bulnes2014May,Croy2016Apr,Watanabe2017Aug,Hino2020Jul} have been for the most part carried out on a very general level in two respects. First of all, these works do not necessarily distinguish between charge or energy currents, and rather focus on the combined heat currents, or even completely generic "place-holder" quantities without specification. Secondly, the additional step to derive the explicit FRR for specific cumulants has been omitted. For this reason, the mechanism underlying recently discovered deviations of the FRR in the presence of time-dependent driving compared to known stationary FRR in systems with strong many-body correlations~\cite{Riwar2013May,Dittmann2018Sep}, remained unclear so far. Furthermore, we believe that this lack of specificity resulted in an incomplete understanding of the importance of many-body interactions in driven systems, first hinted at in
Ref.~\cite{Riwar2013May}.

In the present paper, we set up fluctuation relations and the resulting FRR for interacting quantum pumps, based on the framework of full-counting statistics (FCS) for open quantum systems. We analyze in detail the resulting generalizations of the fluctuation dissipation theorem for the transported charge current and charge current noise. We focus in particular on adiabatic pumping, which is first and foremost of fundamental interest due to the geometric properties of the currents~\cite{Berry1984Mar,Uhlmann1986Oct,Ning1992Oct,Landsberg1992Aug,Brouwer1998Oct,Avron2001Nov,Sarandy2005Jan,Sinitsyn2007Nov,Sinitsyn2007Feb,Yoshii2013Dec,Pluecker2017Apr,Pluecker2017Nov,Nakajima2015Nov,Watanabe2017Aug}. Quantum pumps are furthermore of interest as controlled sources of a quantized current, see Ref.~\cite{Pekola2013Oct} (and references within).

Most importantly, we find in our analysis that for pumping it is in general impossible to formulate FRR for charge transport only, without taking into account the energy input from the pump.
Moreover, we can show that it is the specific geometric properties of the charge and energy transport, which are at the physical origin of deviations from stationary FRR. Namely, the explicit time-dependence of the eigenenergies of the quantum system breaks a global symmetry in the energy counting fields of the cumulant generating functions. As a consequence, the total energy input cannot be eliminated in the geometric response. Specifically, we find that the stationary fluctuation-dissipation theorem is violated due to the non-linear response of this pumping energy input.
While heat transport and heat current fluctuations have been studied in time-dependently driven systems previously~\cite{Battista2014Aug,Battista2014Dec,Moskalets2014May,Moskalets14Err,TerrenAlonso2019Mar,Popovic2020Aug}, this intricate connection between charge and energy transport had to our knowledge remained unknown so far.

Interestingly, we find in addition that the nonlinear-response character of the correction term has an important consequence: deviations from the standard stationary FRR occur only in the presence of many-body interactions.
This, in particular, opens up the possibility to detect correlation effects via deviations from the standard FRR. Finally, we illustrate our general results at the specific example of a driven quantum dot with Coulomb interactions.

The paper is organized as follows. In Sec.~\ref{sec_FCS_weakly_coupled} we review the FCS framework for weakly coupled quantum systems and develop a systematic expansion for slowly driven pumps. Based on this framework, we derive a set of FRR in Sec.~\ref{sec_FR_pumping}, and explore the importance of many-body interactions. Finally, in Sec.~\ref{sec_interactions} we illustrate the derived FRR at the explicit model of a single-level interacting quantum dot and discuss their measurability.

\section{FCS formalism for weakly coupled quantum pumps}\label{sec_FCS_weakly_coupled}

We study the FCS of particle and energy currents through small quantum systems in the regime of weak  tunnel coupling to the reservoirs~\cite{Bagrets2003Feb} including a time-dependent drive~\cite{Croy2016Apr}. In this section we review this framework, and state explicitly the assumptions under which it is valid. We then consistently expand the FCS in orders of the driving frequency, using a similar technique as for closed systems~\cite{Mostafazadeh1997Mar}. For the first order, the adiabatic-response (pumping) contribution, we recover the Sinitsyn-Nemenmann geometric phase~\cite{Sinitsyn2007Nov,Sinitsyn2007Nov}. In fact, the Sinitsyn-Nemenmann phase can be regarded as a geometric phase of the Landsberg type~\cite{Landsberg1992Aug}, where the necessary symmetry stems from the fact that the average pumped current is invariant with respect to a continuous recalibration of the charge meter, see Refs.~\cite{Pluecker2017Apr,Pluecker2017Nov}. The geometric nature of the pumping FCS is crucial for the properties of the cumulant generating function discussed in Sec.~\ref{sec_FR_pumping}.

\subsection{Dynamics of the model system}

We are interested in time-dependently driven quantum systems, exchanging particles and energy with reservoirs. We make the simplifying assumption that the tunnel coupling between the system and the reservoirs is weak, which is standard in the field~\cite{Esposito2006Apr,Andrieux2007Feb,Esposito2007Sep,Esposito2009Dec,Utsumi2010Mar,Ren2010Apr,Campisi2011Jul,Ansari2015Mar,Cuetara2015May,Altaner2016Oct,Benito2016Nov}. To study the effects of adiabatic pumping, we include a periodic driving of the system, which is slow with respect to the decay time of reservoir correlations. We will quantify these conditions in detail, in Sec.~\ref{sec_interactions}.

We describe the dynamics of the system, using its reduced density matrix, represented in vector form $|P)$, which is obtained from the full density matrix by tracing out the reservoir degrees of freedom. 
The reduced density matrix fulfills a master equation
\begin{equation}\label{eq_master_eq}
|\dot{P})=\mathbf{W}(t)|P)\ .
\end{equation}
Here, $\mathbf{W}(t)$ is a kernel, carrying all the transition rates of the stochastic processes that change the system state $|P)$. We here assume that the dynamics of the occupation probabilities of the system states does not couple to the dynamics of coherent superpositions, such that only the diagonal elements of the reduced density matrix matter\footnote{This widely used assumption is in particular valid for the driven single-level quantum dot, treated in Sec.~\ref{sec_interactions}. A generalization to model systems where coherent dynamics are important can be done along the lines of e.g.~\cite{Wunsch2005Nov} or~\cite{Braun2004Nov}}. In the weak coupling limit, which we are considering here, the rates can be computed through Fermi's Golden rule of the \textit{frozen} system where the time-dependence enters parametrically~\cite{Cavaliere2009Sep}.

Through the eigendecomposition of the kernel $\mathbf{W}$, we can understand the dynamics of the quantum system. Namely, the kernel can be decomposed into the form $\mathbf{W}(t)=\sum_k \lambda_k(t) |k(t))(k(t)|$, where the $|k)$ and $(k|$ are the right and left eigenvectors of $\mathbf{W}$, belonging to the eigenvalue $\lambda_k$. The notation is chosen such that any $|a)$ mathematically represents an operator, cast into vector form, whereas any $(a|$ represents a map from an operator to a scalar~\cite{Saptsov2012Dec,Saptsov2014Jul,Schulenborg2014May}. There is a zero mode $k=0$, which expresses that for each $t$ there exists a unique stationary state $|0)$ with $\lambda_0=0$. The corresponding (dual) left eigenvector $(0|$ is the trace operator, expressing the trace preserving property of $\mathbf{W}(t)$. The other eigenvalues and eigenmodes do not play a role in the long-time FCS of this slowly driven system, see Eq.~(40) of Ref.~\cite{Pluecker2017Nov}.

Importantly, the system is coupled to several reservoirs, enumerated with $\alpha$. In the weak coupling limit, the influence from the different reservoirs is additive $\mathbf{W}=\sum_\alpha \mathbf{W}_\alpha$. We assume microreversibility and local equilibrium for each reservoir, such that kernels have the symmetry~\cite{Esposito2009Dec}\footnote{A generalization of the symmetries of the kernel to kernels connecting diagonal and off-diagonal elements of the density matrix can be envisaged along the lines of Ref.~\cite{Ansari2015Mar}}
\begin{equation}\label{eq_kernel_symmetries}
\mathbf{W}_\alpha^T(t)=e^{\beta_{\alpha}\left[\mu_{\alpha}\mathbf{n}-\mathbf{e}(t)\right]}\mathbf{W}_{\alpha}(t)e^{-\beta_{\alpha}\left[\mu_{\alpha}\mathbf{n}-\mathbf{e}(t)\right]} \ .
\end{equation}
where $\beta_\alpha=1/(k_\text{B}T_\alpha)$ and $\mu_\alpha$ are, respectively, the inverse temperature and chemical potential of reservoir $\alpha$. The particle number and energy superoperators associated to the local quantum system, $\mathbf{n}=\frac{1}{2}\{\widehat{n},\cdot\}$ and $\mathbf{e}=\frac{1}{2}\{\widehat{\epsilon},\cdot\}$, are defined via the anticommutators of the local particle number and energy operators, $\widehat{n}$ and $\widehat{\epsilon}$, respectively. They will be explicitly defined, when considering an explicit model system, see Sec.~\ref{sec_interactions}.

We use the following notation throughout this work. The hat designates operators, which, when cast into vector form (in the superoperator context), are written in the round bra notation, $|\ldots)$. Superoperators are written in bold font, whereas regular fonts are used for scalars.

Finally, let us note that in principle, the chemical potential and temperature gradients may be time-dependent, too, since Eq.~\eqref{eq_kernel_symmetries} is a time-local symmetry. However, in the remainder of this work, we keep $\mu_\alpha$ and $\beta_\alpha$ constant in time in order to be able to clearly separate the response due to driving and the response due to chemical potential or temperature gradients. We will consider a time-dependent driving of the local system parameters, and the coupling amplitudes to the reservoirs, see Sec.~\ref{sec_interactions} for a concrete example.

\subsection{Full counting statistics}

We are not only interested in the mean dynamics of the system, but also in the FCS. Through integrating out the reservoir degrees of freedom, one has in principle lost all information about the transport statistics. However, one can keep track of the information of particle and energy transport by supplementing the kernels with counting fields (see Ref.~\cite{Esposito2009Dec} for a review),
\begin{equation}\label{eq_W_chi_xi}
\mathbf{W}(\{\chi_\alpha,\xi_\alpha\},t)=\sum_\alpha e^{-i\mathbf{e}(t)\xi_{\alpha}}e^{-i\mathbf{n}\chi_{\alpha}}\mathbf{W}_{\alpha}(t)e^{i\mathbf{n}\chi_{\alpha}}e^{i\mathbf{e}(t)\xi_{\alpha}}\ .
\end{equation}
The counting fields $\chi_\alpha$ and $\xi_\alpha$ keep track of the number of particles and the energy that enter reservoir $\alpha$. Let us stress again, our main interest is in the charge currents, counted by $\chi_\alpha$. However, as we will show, the FRR for the charge current cannot be formulated and interpreted unless the energy currents are accounted for as well, such that we have to keep $\xi_\alpha$.

Also for the kernel including counting fields, microreversibility imposes a symmetry, similar to Eq.~\eqref{eq_kernel_symmetries}, which can be expressed in terms of the counting fields through
\begin{equation}\label{eq_W_symm}
\mathbf{W}^T(\{\chi_\alpha,\xi_\alpha\},t)=\mathbf{W}(\{i\beta_\alpha\mu_\alpha-\chi_\alpha,-i\beta_\alpha-\xi_\alpha\},t)\ .
\end{equation}
This symmetry is of central importance for the formulation of fluctuation relations out of equilibrium~\cite{Esposito2006Apr,Andrieux2007Feb,Esposito2007Sep,Esposito2009Dec,Utsumi2010Mar,Campisi2011Jul,Lippiello2014Apr,Borrelli2015Jan,Cuetara2015May,Altaner2016Oct,Benito2016Nov}. 

In order to obtain the particle and energy transport statistics, one needs to construct a so-called cumulant-generating function. 
One therefore starts from an arbitrary initial state $|P_0)$ at time $t_\text{in}$ and then switches on the counting fields to measure the FCS until a certain time $\tau$. Tracing over the remaining system degrees of freedom through application of $(0|$, one finds the cumulant generating function, $\mathcal{F}(\{\chi_\alpha,\xi_\alpha\},\tau,t_\text{in})$, from the evolution of the system in presence of the counting fields
\begin{equation}\label{eq_cumulant_general}
e^{\mathcal{F}(\{\chi_\alpha,\xi_\alpha\},\tau,t_\text{in})(\tau-t_\text{in})}=(0|\Pi(\{\chi_\alpha,\xi_\alpha\},\tau,t_\text{in})|P_0)\ .
\end{equation}
Here, we have introduced the propagator of the open system
\begin{equation}
\Pi(\{\chi_\alpha,\xi_\alpha\},\tau,t_\text{in})=\mathcal{T}e^{\int_{t_\text{in}}^\tau dt \mathbf{W}(\{\chi_\alpha,\xi_\alpha\},t)}\ .
\end{equation}
The cumulants of the charge- and energy currents can now be computed by differentiating with respect to the counting fields of interest, and subsequently setting all counting fields to zero. For the remainder of this paper, we focus on the limit of very long measurement times $(\tau-t_\text{in})/t_\text{typ}\rightarrow\infty$, namely when $\tau-t_\text{in}$ is much larger than typical system time scales $t_\text{typ}$ given by the inverse of kernel eigenvalues $\lambda_k\neq0$. Then $\mathcal{F}$ neither depends on $\tau$ nor $t_\text{in}$ and provides the zero-frequency cumulants. In this case, the FCS does no longer depend on the initial state $|P_0)$, since for $\tau-t_\text{in}\rightarrow\infty$, only the stationary state $|0)$ contributes to the transport statistics. Without loss of generality, we therefore set our initial state to be the stationary state in absence of the transport counting $|P_0)=|0(0))$.
The zero-frequency cumulants of interest in the present paper, namely the average charge current into reservoir $\alpha$ and the related current-current correlations, are then defined as
\begin{align}\nonumber
I_\alpha &\equiv I_\alpha(\tau-t_\text{in}\to\infty)\\\label{eq_def_I} & = -i \lim_{\tau-t_\text{in}\to\infty}\left[\partial_{\chi_\alpha} \left.\mathcal{F}\right|_{\{\chi_\alpha,\xi_\alpha\}\rightarrow 0}\right]\ ,
\end{align}
as well as
\begin{align}\nonumber
S_{\alpha\gamma} &\equiv S_{\alpha\gamma}(\tau-t_\text{in}\to\infty)\\ \label{eq_def_S} & = - \lim_{\tau-t_\text{in}\to\infty}\left[\partial_{\chi_\alpha}\partial_{\chi_\gamma} \left.\mathcal{F}\right|_{\{\chi_\alpha,\xi_\alpha\}\rightarrow 0}\right]\ .
\end{align}
Importantly, and as foreshadowed already, also the energy current into reservoir $\alpha$ is found to play an important role in the present paper; it is defined via a derivative with respect to the energy counting field $\xi_\alpha$,
\begin{align}\nonumber
I_{E,\alpha}&\equiv I_{E,\alpha}(\tau-t_\text{in}\to\infty)\\\label{eq_I_E_definition} & = -i \lim_{\tau-t_\text{in}\to\infty}\left[\partial_{\xi_\alpha} \left.\mathcal{F}\right|_{\{\chi_\alpha,\xi_\alpha\}\rightarrow 0}\right].
\end{align}

\subsection{Adiabatic expansion}

In order to find explicit expressions for the cumulant generating function, we now focus on the limit of slow driving as previously considered in~\cite{Sinitsyn2007Feb,Hino2020Jul}. This means that the time-scale related to the inverse frequency of the driving $\tau_0=2\pi/\Omega$, is large with respect to the time scale on which the system states vary. In this adiabatic limit, the cumulant generating function $\mathcal{F}$ can be expanded in orders of the driving parameters, see Appendix~\ref{appendix_adiabatic}.  We evaluate it up to first order in the small driving parameter, $\mathcal{F}\approx\mathcal{F}^{(0)}+\mathcal{F}^{(1)}$. The zeroth-order, instantaneous contribution is given by\footnote{Note that for simplicity of notation, we omitted the explicit time arguments in the eigenvalues and eigenvectors of $W$.}
\begin{equation}\label{eq_F_inst}
\mathcal{F}^{(0)}(\{\chi_\alpha,\xi_\alpha\})=\int_0^{\tau_0}\frac{dt}{\tau_0}\lambda_0(\{\chi_\alpha,\xi_\alpha\})\ .
\end{equation}
This is just a time-averaged version of the FCS which appear for a system without time-dependent driving. On top of that, we find the pumping contribution
\begin{equation}\label{eq_pumping_FCS}
\begin{split}
\mathcal{F}^{(1)}(\{\chi_\alpha,\xi_\alpha\})=-\int_0^{\tau_0}\frac{dt}{\tau_0}(0(\{\chi_\alpha,\xi_\alpha\})|\partial_t|0(\{\chi_\alpha,\xi_\alpha\}))\ .
\end{split}
\end{equation}
This reproduces the result derived in Ref.~\cite{Sinitsyn2007Feb} and shows the clearly geometric property of the adiabatic pumping transport statistics. This has important consequences: first of all, it is the properties of the \textit{eigenvectors} and not only of the eigenvalues that enter here. Second, due to the time-dependent driving of system energies $\mathbf{e}(t)$, it can already be expected that the time-derivative will lead to terms involving the energy counting field, see also Eq.~(\ref{eq_W_chi_xi}).

\section{Charge current response relations and interaction effects}\label{sec_FR_pumping}

Our main goal is the derivation and interpretation of the FRR for charge currents using the symmetries due to microreversibility~\cite{Esposito2009Dec} of the cumulant generating function in the counting fields. Importantly, we show, that the geometric nature of the pumping contribution forbids in general that the charge FRR can be formulated without the appearance of an additional contribution, which is associated to the energy current provided by the external pumping fields. We find this extra term to be a consequence of a drive-induced breaking of gauge invariance with respect to the energy counting fields $\xi_\alpha$. We then make the additional surprising observation that the presence or absence of the additional energy dissipation term hinges on the presence or absence, respectively, of many-body interactions. Interestingly, while distinct in terms of the physical origin, this fact falls in line with various other interaction-induced pumping effects, see Refs.~\cite{Splettstoesser2006Aug,Reckermann2010Jun}.

\subsection{Fluctuation relations}\label{sec_symmetries_and_gauge}
Importantly, due to the above introduced symmetries of $\mathbf{W}(\{\chi_\alpha,\xi_\alpha\})$ [see Eq.~\eqref{eq_W_symm}], we can derive fluctuation relations for $\mathcal{F}$. Namely we find that the instantaneous cumulant generating function satisfies the same relations as for a system without driving,
\begin{equation}\label{eq_cgf_0}
\mathcal{F}^{(0)}(\{\chi_\alpha,\xi_\alpha\})=\mathcal{F}^{(0)}(\{i\beta_\alpha\mu_\alpha-\chi_\alpha,-i\beta_\alpha-\xi_\alpha\})\ ,
\end{equation}
whereas the pumping contribution satisfies
\begin{equation}\label{eq_cgf_1}
\mathcal{F}^{(1)}(\{\chi_\alpha,\xi_\alpha\})=-\mathcal{F}^{(1)}(\{i\beta_\alpha\mu_\alpha-\chi_\alpha,-i\beta_\alpha-\xi_\alpha\})\ .
\end{equation}
The minus sign for the pumping cumulant generating function reflects the fact that in order to satisfy microreversibility, the direction of pumping transport has to be inverted, too.
The result shown in Eqs.~\eqref{eq_cgf_0} and~\eqref{eq_cgf_1} has in another form been found previously~\cite{Hino2020Jul} (i.e., in the form of a generic counting field, which could in principle encompass either charge or energy, or another observable quantity). Contrary to Ref.~\cite{Hino2020Jul} however, we explicitly introduce the separate counting fields of charge and energy currents. It is this explicit distinction, which allows a careful derivation and interpretation of charge current FRR, as we will present them in the following.

\subsection{Gauge transformations and their relationship to current conservation}

As already announced, the first crucial step is to consider global gauge transformations of the cumulant generating function with respect to the counting fields.
Both the instantaneous ($i=0$) and the pumping contribution ($i=1$) to the cumulant generating function, $\mathcal{F}$, are invariant with respect to global shifts of the \textit{particle} counting fields,
\begin{equation}\label{eq_charge_F}
\mathcal{F}^{(i)}(\{\chi_\alpha+\delta\chi,\xi_\alpha\})=\mathcal{F}^{(i)}(\{\chi_\alpha,\xi_\alpha\})\ .
\end{equation} 
This invariance reflects the conservation of charge currents, which is valid irrespective of the presence or absence of a time-dependent driving. When applied on the level of the individual cumulants, we can derive the important and well-known identities 
\begin{subequations}
\begin{align}\label{eq_current_conservation}
    \sum_\alpha I_\alpha^{(i)}&=0\\
    \sum_\alpha S_{\alpha\gamma}^{(i)}=\sum_\gamma S_{\alpha\gamma}^{(i)}&=0.
\end{align}
\end{subequations}
However, the instantaneous and pumping contributions behave fundamentally differently from Eq.~(\ref{eq_charge_F}) with respect to a similar gauge in the energy counting fields, a fact which will be central for the rest of the discussion. While the instantaneous contribution exhibits a similar gauge invariance for $\xi$,
\begin{equation}\label{eq_en_F_0}
\mathcal{F}^{(0)}(\{\chi_\alpha,\xi_\alpha+\delta\xi\})=\mathcal{F}^{(0)}(\{\chi_\alpha,\xi_\alpha\})\ ,
\end{equation}
the pumping correction does not. On the contrary, here, we receive an extra term
\begin{equation}\label{eq_xi_shift}
\mathcal{F}^{(1)}(\{\chi_\alpha,\xi_\alpha+\delta\xi\})=\mathcal{F}^{(1)}(\{\chi_\alpha,\xi_\alpha\})-i\delta\xi\,\mathcal{Q}^{(1)}(\{\chi_\alpha,\xi_\alpha\})\ ,
\end{equation}
where we defined
\begin{equation}
\mathcal{Q}^{(1)}(\{\chi_\alpha,\xi_\alpha\})=-\int_0^{\tau_0}\frac{dt}{\tau_0}(0(\{\chi_\alpha,\xi_\alpha\})|\dot{\mathbf{e}}|0(\{\chi_\alpha,\xi_\alpha\}))\ .
\end{equation}
Formally, the origin of this different gauge behaviour, Eqs. (\ref{eq_charge_F}), (\ref{eq_en_F_0}) and (\ref{eq_xi_shift}), can be found when considering the kernel $\mathbf{W}(\{\chi_\alpha,\xi_\alpha\})$. Constant global shifts of the counting fields result in unitary transformations of the kernel, $\mathbf{W}(\{\chi_\alpha+\delta\chi,\xi_\alpha\})=e^{-i\mathbf{n}\delta\chi}\mathbf{W}(\{\chi_\alpha,\xi_\alpha\})e^{i\mathbf{n}\delta\chi}$ respectively $\mathbf{W}(\{\chi_\alpha,\xi_\alpha+\delta\xi\})=e^{-i\mathbf{e}(t)\delta\xi}\mathbf{W}(\{\chi_\alpha,\xi_\alpha\})e^{i\mathbf{e}(t)\delta\xi}$. Consequently, the \textit{eigenvalues} of $\mathbf{W}$ remain unchanged, hence the gauge invariance of $\mathcal{F}^{(0)}$. The \textit{eigenvectors} however, are transformed through the unitary super-operators $\exp(\pm i\mathbf{n}\delta\chi)$ and $\exp(\pm i \mathbf{e}(t)\delta\xi)$. Crucially, while a global shift in the particle counting field results in a time-independent transformation, the shift in the energy counting fields produces a \textit{time-dependent} transformation. Hence, because of the geometric form of $\mathcal{F}^{(1)}$, see Eq.~(\ref{eq_pumping_FCS}), it is easy to realize that the latter shift can in general not be eliminated, but results in the extra term on the right hand side of Eq.~(\ref{eq_xi_shift}).

Moreover, the breaking of this global symmetry can also be understood in physical terms. Namely, starting from Eq.~\eqref{eq_en_F_0}, and solving for $\mathcal{Q}$, it is possible to show that
\begin{equation}\label{eq_G_vs_F}
-i\mathcal{Q}^{(1)}=\sum_\gamma \partial_{\xi_\gamma} \mathcal{F}^{(1)}\ .    
\end{equation}
Consequently, when putting all counting fields to zero, we can directly relate this function to the energy current provided by the external pumping fields,
\begin{equation}\label{eq_G_IE}
\mathcal{Q}^{(1)}(\{0,0\})=-\int_0^{\tau_0}\frac{dt}{\tau_0}(0|\dot{\mathbf{e}}|0)=-\sum_\alpha I_{E,\alpha}^{(1)}\equiv I_{E,\text{pump}}^{(1)} \ .
\end{equation}
This reflects the very simple fact that energy is in general not conserved in the presence of an external drive, and can instead be pumped into the system. Charge conservation on the other hand is guaranteed even in the presence of the drive, such that no similar term arises for shifts in $\chi$.

Let us now proceed and study the FRR for the pumping \textit{charge} currents. We therefore have to determine the degree to which charge and energy counting fields can be disentangled in the presence of driving. The above derived gauge considerations are instrumental for this. From now on, we set $\beta_\alpha=\beta$ to avoid that the effect of driving is obscured.  For the instantaneous contribution, Eq.~\eqref{eq_cgf_0}, the energy counting fields can then just be set to zero (no counting of energy currents), and the global shift $-i\beta$ appearing in the energy counting field argument can be gauged away [due to Eq.~\eqref{eq_en_F_0}]. This simply gives rise to
\begin{equation}\label{eq_FR_F0}
\mathcal{F}^{(0)}(\{\chi_\alpha\})=\mathcal{F}^{(0)}(\{i\beta\mu_\alpha-\chi_\alpha\})\ .
\end{equation}
This is the time-averaged version of the well-known fluctuation relations for time-independent systems~\cite{Esposito2006Apr,Andrieux2007Feb}  (see also Eq.~(\ref{eq_F_inst})). Based on the above discussion, it is obvious that we cannot achieve the same "charge current-only" response relations for the pumping contribution, Eq.~\eqref{eq_cgf_1}. Namely, one cannot get rid of the pumping energy input $I_{E,\text{pump}}$, and instead receives
\begin{equation}\label{eq_FR_F1_G}
\mathcal{F}^{(1)}(\{\chi_\alpha\})=-\mathcal{F}^{(1)}(\{i\beta\mu_\alpha-\chi_\alpha\})+\beta\,\mathcal{Q}^{(1)}(\{i\beta\mu_\alpha-\chi_\alpha\})\ .
\end{equation}
We note that Eq.~\eqref{eq_FR_F1_G} automatically implies that 
\begin{equation}\label{eq_FR_G1}
\mathcal{Q}^{(1)}(\{\chi_\alpha\})=\mathcal{Q}^{(1)}(\{i\beta\mu_\alpha-\chi_\alpha\})\ .
\end{equation}
such that, remarkably $\mathcal{Q}^{(1)}$ itself \textit{does} satisfy a fluctuation relation quite similar to the instantaneous $\mathcal{F}^{(0)}$.

Equation \eqref{eq_FR_F1_G} is one of the central results of this paper. We conclude that it is in general impossible to derive FRR for the charge pumping contribution, which involve only charge currents. The energy currents always enter due to the presence of the term $\mathcal{Q}^{(1)}$. We will see in the next section, that this contribution due to the pumping energy input is omnipresent when expressing the pumping fluctuation relations in terms of the individual cumulants, that is, the response relations.

\subsection{FRR for charge pumping}\label{sec_FR_for_charge_cumulants}

The cumulant generating functions can now be expanded in terms of the counting fields and  of the gradients in chemical potentials, in order to derive a resulting hierarchy of equations relating cumulants of different orders, yielding FRR. The derivation is detailed in Appendix~\ref{appendix_straightforward}.

In fact, the FRR for both instantaneous and first order in pumping can be expressed in a very compact form when introducing the total power provided by the external pumping field and the applied biases
\begin{equation}\label{eq_J_pump_def}
J_\text{pump}\equiv -\sum_\alpha\left(I_{E,\alpha}-\mu_\alpha I_\alpha\right)\equiv I_{E,\text{pump}}+\sum_\alpha\mu_\alpha I_\alpha\ .
\end{equation}
Then, we find both for the instantaneous ($i=0$) and for the adiabatic-response contribution ($i=1$),
\begin{align}\label{eq_FRR_J1}
0&=\left.\frac{\partial J_\text{pump}^{(i)}}{\partial\mu_\alpha}\right|_{\{\mu_\alpha\}\rightarrow\mu} \\ \label{eq_FRR_J2}
\left.S_{\alpha\gamma}^{(i)}\right|_{\{\mu_\alpha\}\rightarrow\mu}&=k_\text{B}T\left.\frac{\partial^2 J^{(i)}_\text{pump}}{\partial\mu_\alpha\partial\mu_\gamma}\right|_{\{\mu_\alpha\}\rightarrow\mu}\ .
\end{align}
However, when expressed in terms of the power, the subtly different behavior of the instantanteous and pumping fluctuation relations is not clearly visible. We therefore from now on distinguish explicitly charge and energy currents, and discuss the resulting relations in detail.

For the instantaneous order, we get a time-averaged version of the response relations found also in Ref.~\cite{Esposito2006Apr,Andrieux07}. The lowest order relations read
\begin{align}\label{eq_FR_I0}
\left.I_{\alpha}^{(0)}\right|_{\{\mu_\alpha\}\rightarrow\mu}&=0\\ \label{eq_FR_S0}
\left.S_{\alpha\gamma}^{(0)}\right|_{\{\mu_\alpha\}\rightarrow\mu}&=k_\text{B}T\left.\left(\frac{\partial I_\alpha^{(0)}}{\partial\mu_\gamma}+\frac{\partial I_\gamma^{(0)}}{\partial\mu_\alpha}\right)\right|_{\{\mu_\alpha\}\rightarrow\mu}
\end{align}
The first relation states that the time-averaged instantaneous currents must be zero in the absence of gradients in the chemical potentials, as expected. The second relation is a time-averaged version of the famous fluctuation-dissipation theorem. 

As for the adiabatic-response, i.e. the pumping contribution, we find a first relation of the form
\begin{equation}\label{eq_FR_zeroth}
0=\left.I_{E,\text{pump}}^{(1)}\right|_{\{\mu_\alpha\}\rightarrow\mu}\ ,
\end{equation}
stating very simply that the pumping energy input is zero in the absence of chemical potential differences. This relation has no equivalent (or would be trivial) in the instantaneous order. It can be interpreted as follows. While one can induce a pumping current in the absence of chemical potential gradients (see also the FRR below), the electrons will be shuffled from one reservoir to the other, without putting in or extracting work. In addition, we find
\begin{align}\label{eq_FR_I1}
\left.I_{\alpha}^{(1)}\right|_{\{\mu_\alpha\}\rightarrow\mu}&=\left.\frac{\partial I_{E,\text{dis}}^{(1)}}{\partial\mu_\alpha}\right|_{\{\mu_\alpha\}\rightarrow\mu} \\ \label{eq_FR_S1}
\left.S_{\alpha\gamma}^{(1)}\right|_{\{\mu_\alpha\}\rightarrow\mu}&=\\
&k_\text{B}T\left.\left(\frac{\partial I_\alpha^{(1)}}{\partial\mu_\gamma}+\frac{\partial I_\gamma^{(1)}}{\partial\mu_\alpha}-\frac{\partial^2 I_{E,\text{pump}}^{(1)}}{\partial\mu_\alpha\partial\mu_\gamma}\right)\right|_{\{\mu_\alpha\}\rightarrow\mu}  \nonumber\ .
\end{align}
Here, we see explicitly, what we have already indicated in the previous section, in Eq.~\eqref{eq_FR_F1_G}, namely that the pumping energy input necessarily appears when attempting to formulate relations between different charge current cumulants. This seems to have been overlooked so far. While it is a well-known result (in many different physical regimes) that the pumping current can be non-zero even in the absence of a voltage bias~\cite{Buttiker1994Mar,Brouwer2001Mar,Avron2004Aug,Splettstoesser2005Dec,Splettstoesser2006Aug} we here show, that this nonzero pumping current can be related to the linear response of the energy current provided by the external pumping fields, $I_{E,\text{pump}}$.

In Eq.~\eqref{eq_FR_S1}, we recover a second FRR for pumping. Namely, while the instantaneous relation for the current noise, Eq.~\eqref{eq_FR_S0}, satisfies a stationary (equilibrium) fluctuation dissipation theorem, this simple form cannot be extended to the pumping noise. Instead, it receives a correction due to the nonlinear response of $I_{E,\text{pump}}$. A deviation from the fluctuation-dissipation theorem for the pumping noise has already been noted by us in Ref.~\cite{Riwar2013May} for a concrete single-level quantum dot model. In other works it has been implied that the stationary FDT does not extend to first order due to the geometric nature of the pumping cumulant generating function~\cite{Ren2010Apr,Hino2020Jul}. However, in neither of these works the deviation from the stationary FDT has been explicitly identified or interpreted. Here, we find its explicit physical meaning  as the nonlinear response of the pumping energy input.

\subsection{The importance of charge conservation and many-body interactions}

In this section, we show under which conditions the deviations from the standard FRR stemming from nonlinearities in the pumping energy input play a role.

The term, which we want to analyze in more detail here is the nonlinearity in the pumping energy input
\begin{align}
    \left.\frac{\partial^2 I_{E,\text{pump}}^{(1)}}{\partial\mu_\alpha\partial\mu_\gamma}\right|_{\{\mu_\alpha\}\rightarrow\mu}\ .
\end{align}
We start by considering the cross correlations, namely $\alpha\neq\gamma$. One immediately notices then that this term is nonzero only, if at least one of the summands in $I_{E,\text{pump}}^{(1)}$ depends on the electrochemical potential of \textit{two different} reservoirs. However, it is well known that in systems where many-body interactions are negligible the pumping charge and energy currents can be written in terms of incoming and outgoing effective distributions, where the different electrochemical potentials contribute to different summands, see e.g. Eqs. (9) and (12) of Ref.~\cite{Moskalets2002Nov}.
This results in 
\begin{align}
    \left.\frac{\partial^2 I_{E,\text{pump}}^{(1)}}{\partial\mu_\alpha\partial\mu_\gamma}\right|_{\{\mu_\alpha\}\rightarrow\mu}\xrightarrow{\text{no\ interactions}} 0\ .
\end{align}
 Crucially, we thus find that an equilibrium version of the FRR can be recovered for vanishing interactions. This is a further main result of this work. The deviation from the stationary FDT for the pumping response constitutes a purely interaction-induced effect, which is expected to be directly measurable in a noise or counting experiment. Moreover, this is a significant generalization of the result found in Ref.~\cite{Riwar2013May}, where the recovery of the FRR for the pumping noise was found for the special case of a single-level quantum dot, including only two contacts. Here, the proof encompasses the broad class of generic quantum systems, weakly coupled to an arbitrary number of reservoirs, see Sec.~\ref{sec_FCS_weakly_coupled}.

We now show how the above extends to autocorrelations via current conservation.
Using Eq.~(\ref{eq_current_conservation}), one directly observes that the autocorrelations can be expressed as 
\begin{align}\label{eq_autocross}
    \left.S_{\alpha\alpha}^{(1)}\right|_{\{\mu_\alpha\}\rightarrow\mu}=-\sum_{\gamma\neq\alpha}\left.S_{\alpha\gamma}^{(1)}\right|_{\{\mu_\alpha\}\rightarrow\mu}
\end{align}
The constraint given in Eq.~(\ref{eq_autocross}) can be rewritten using Eq.~(\ref{eq_FR_S1}), leading to
\begin{equation}\label{eq_d_mu_gamma}
\sum_\gamma\left.\frac{\partial I^{(1)}_\alpha}{\partial \mu_\gamma}\right|_{\{\mu_\alpha\}\rightarrow\mu}=\sum_\gamma\left.\frac{\partial^2 I_{E,\text{pump}}^{(1)}}{\partial\mu_\alpha\partial\mu_\gamma}\right|_{\{\mu_\alpha\}\rightarrow\mu}  \ .
\end{equation}
Note, that the left hand side of this equality is the sum over the adiabatic-response corrections to the charge conductances. In a non-driven system, this sum has to always equal zero. However, since the driving fields induce pumping charge currents, the extra term on the right hand side of Eq.~(\ref{eq_d_mu_gamma}) appears, thereby constituting a generalization of the conductance sums for time-dependently driven systems.

Equation~\eqref{eq_d_mu_gamma} can be rewritten as
\begin{equation}\label{eq_d_mu}
\partial_\mu\left(\left.I^{(1)}_\alpha\right|_{\{\mu_\alpha\}\rightarrow\mu}\right)=\partial_\mu\left(\left.\frac{\partial I_{E,\text{pump}}^{(1)}}{\partial\mu_\alpha}\right|_{\{\mu_\alpha\}\rightarrow\mu}\right)  \ .
\end{equation}
This shows that the correction of the sum of conductances can be derived from Eq.~(\ref{eq_FR_I1}), by means of the derivative with respect to $\mu$. Hence, the additional constraints for the auto-correlations are already fixed by the FRR for currents, Eq.~(\ref{eq_FR_I1}).

\section{Explicit model and the role of interactions}\label{sec_interactions}

We have understood on a general level, how pumping modifies the FRR. The microreversibility relation, Eq.~\eqref{eq_W_symm}, corresponds to a gauge transformation, which leads to an extra term related to the energy current provided by the external pumping fields. We now consider an explicit  quantum dot pump model to examine the properties of the fluctuation relations in detail. In particular we focus on how to probe the fluctuation relations in this experimentally relevant system, and illuminate the special role of electron-electron interactions.

\begin{figure}
\centering
\includegraphics[width=0.45\textwidth]{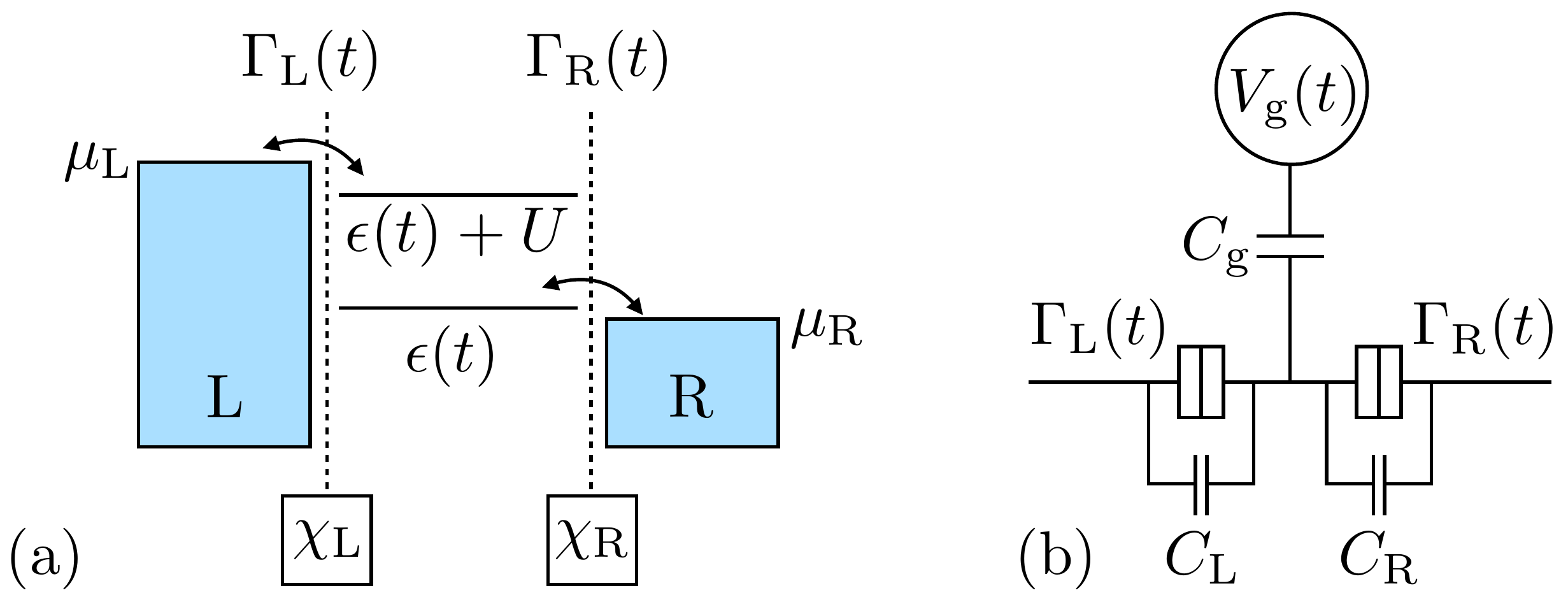}
\caption{(a) Energy diagram of the single-level quantum dot. The quantum dot level has a time-dependent energy $\epsilon(t)$ and is tunnel coupled to a left (L) and right lead (R), with respective tunneling rates $\Gamma_\text{L,R}(t)$. The reservoirs may have a chemical potential difference, $\mu_\text{L}\neq\mu_\text{R}$. The charge counting fields $\chi_\text{L,R}$ are indicated at the tunnel barriers. (b) The circuit picture of the single-level quantum dot, valid when assuming geometric capacitances.}\label{fig_SQD}
\end{figure}

\subsection{FRR for a single-level quantum-dot pump}
We consider a quantum dot model with a single, spin-degenerate level, tunnel-coupled to two reservoirs which may be at a different chemical potential, see Fig.~\ref{fig_SQD}(a). We have analyzed the charge pumping noise of this model in Ref.~\cite{Riwar2013May}, and found that a deviation from the stationary fluctuation dissipation theorem occurs due to the time-dependent driving. Here, we identify the concrete physical mechanism for this deviation based on the pumping fluctuation relations presented in the previous section.

The total Hamiltonian is of the form
\begin{equation}
\widehat{H}=\widehat{H}_\text{QD}(t)+\widehat{H}_\text{T}(t)+\sum_{\alpha=\text{L,R}}\widehat{H}_\alpha\ .
\end{equation}
The local quantum dot Hamiltonian can be expressed as
\begin{equation}\label{eq_H_QD}
\widehat{H}_\text{QD}=\sum_{\sigma} \epsilon(t) \widehat{d}_\sigma^{\dagger}\widehat{d}_\sigma + \frac{U}{2}\widehat{n}\left(\widehat{n}-1\right)\equiv \widehat{\epsilon}\ .
\end{equation}
Here, the operators $\widehat{d}_\sigma^{(\dagger)}$ annihilate (create) an electron with spin $\sigma$ and energy $\epsilon(t)$. The parameter $U$ expresses the magnitude of the onsite Coulomb interaction, which is sensitive to the electron occupation number $\widehat{n}=\sum_\sigma \widehat{d}_\sigma^\dagger \widehat{d}_\sigma$. As indicated, the quantum dot Hamiltonian corresponds to the local energy operator $\widehat{\epsilon}$. Out of these two operators, one can construct the explicit superoperators $\mathbf{n}$ and $\mathbf{e}$, as introduced in Sec.~\ref{sec_FCS_weakly_coupled}. Note that while $U$ may be time-dependent, too (as we will discuss later) we here keep it constant, in order to clearly distinguish between the interacting and non-interacting regimes. The Hamiltonians of the reservoirs are given as
\begin{equation}
\widehat{H}_\alpha =\sum_{k\sigma}\epsilon_k \widehat{c}_{\alpha k\sigma}^\dagger \widehat{c}_{\alpha k\sigma}
\end{equation}
where likewise the operators $\widehat{c}_{\alpha k\sigma}^{(\dagger)}$ annihilate (create) an electron in reservoir $\alpha$, with momentum $k$ and spin $\sigma$, as well as energy $\epsilon_k$.
Finally, the tunneling Hamiltonian is
\begin{equation}
\widehat{H}_\text{T}(t)=\sum_{\alpha k\sigma}\gamma_\alpha(t)\widehat{c}_{\alpha k\sigma}^\dagger \widehat{d}_\sigma+\text{h.c.}\ ,
\end{equation}
where $\gamma_\alpha(t)$ denotes the time-dependent tunneling amplitude. All the time-dependent parameters shall be subject to a periodic driving, $x(t+\tau_0)=x(t)$, where the pumping period is $\tau_0=2\pi/\Omega$ with the driving frequency $\Omega$. To simplify the notation, we will from now on omit the brackets $(t)$ for these parameters, and for any further quantities depending on them.

We now present the rate equation of the quantum dot system, in presence of the tunnel coupling. In the sequential tunneling limit, valid for small tunnel couplings, $\Gamma_\alpha\ll k_\text{B}T$, with $\Gamma_\alpha=2\pi\rho_\alpha|\gamma_\alpha|^2$, the dynamics of the system are of the form of Eq.~\eqref{eq_master_eq} introduced in Sec.~\ref{sec_FCS_weakly_coupled}. For the here considered system, the reduced density matrix in vector form reads $|P)=(P_0,P_\uparrow,P_\downarrow,P_2)^T$, where only the diagonal elements are relevant\footnote{For the model considered here, this is valid as long as the contacts are normal metals, guaranteeing charge and spin conservation.}. They contain the occupation probabilities of the quantum dot with the probability of the quantum dot being empty $P_0$, singly occupied with either an $\uparrow$ or $\downarrow$ electron $P_{\uparrow,\downarrow}$, or doubly occupied $P_2$.
In the following, we focus on a regime of adiabatic driving\footnote{Depending on the driving amplitudes, this condition should be generalized to $\delta \varepsilon \Omega/\Gamma k_\text{B}T\ll1$ with the amplitude of the time-dependent energy level $\delta\varepsilon$, see Ref.~\cite{Reckermann2010Jun}}, $\Omega\ll \Gamma_\alpha$, where the expansion presented in Appendix~\ref{appendix_adiabatic} applies\footnote{In principle, this allows us to consider the dynamics even for driving faster than the tunneling dynamics, as long as the driving occurs on time scales slower than $(k_\text{B}T)^{-1}$.} and where we compute the dynamics following Refs.~\cite{Splettstoesser2006Aug,Cavaliere2009Sep}.

The kernel describing the time evolution can be given as $\mathbf{W}=\sum_{\alpha=\text{L,R}}\mathbf{W}_\alpha$ and,
\begin{equation}
\mathbf{W}_\alpha=\Gamma_{\alpha}\left(\begin{array}{cccc}
-2f_{\alpha} & \overline{f}_{\alpha} & \overline{f}_{\alpha} & 0\\
f_{\alpha} & -\overline{f}_{\alpha}-f_{\alpha}^{U} & 0 & \overline{f}_{\alpha}^{U}\\
f_{\alpha} & 0 & -\overline{f}_{\alpha}-f_{\alpha}^{U} & \overline{f}_{\alpha}^{U}\\
0 & f_{\alpha}^{U} & f_{\alpha}^{U} & -2\overline{f}_{\alpha}^{U}
\end{array}\right)
\end{equation}
with $f_\alpha=f(\epsilon-\mu_\alpha)$, $f_\alpha^U=f(\epsilon+U-\mu_\alpha)$, and $\overline{f}=1-f$, where $f(E)=1/(e^{E/k_\text{B}T}+1)$ is the Fermi function.
Likewise, we can write the particle number and energy superoperators as $\mathbf{n}=\text{diag}(0,1,1,2)$ and $\mathbf{e}=\text{diag}(0,\epsilon,\epsilon,2\epsilon+U)$.
With the above ingredients, we can now construct the kernel including the particle and energy current counting fields, according to Eq.~\eqref{eq_W_chi_xi}. Thus we find the FCS of a quantum dot pump, and can therefore test the fluctuation relations elaborated in the previous sections. In particular, the kernels $\mathbf{W}_\alpha$ fulfill the symmetry relation of Eq.~(\ref{eq_kernel_symmetries}), and consequently Eq.~\eqref{eq_W_chi_xi} follows from that.

The analytic expressions for the pumping current and pumping noise for this single-level quantum-dot pump have already been computed in Ref.~\cite{Riwar2013May}. For the purpose of this paper, we need the expression for the current for arbitrary chemical potentials (because we will have to compute the conductance below), whereas the noise is only required at $\mu_\text{L}=\mu_\text{R}=\mu$. The respective expressions are
\begin{align}
\left.I_\text{L}^{(1)}\right|_{\{\mu_\alpha\}\rightarrow\mu}& =-\frac{1}{2}\int_{0}^{\tau_0}\frac{dt}{\tau_0}\frac{\lambda_\text{c,L}-\lambda_\text{c,R}}{\lambda_\text{c}}\partial_{t}\left\langle \widehat{n}\right\rangle \\ \label{eq_noise_analytic_zero_bias}
\left.S_\text{LR}^{(1)}\right|_{\{\mu_\alpha\}\rightarrow\mu}&=\int_{0}^{\tau_0}\frac{dt}{\tau_0}\frac{\Gamma_{\text{L}}\Gamma_{\text{R}}}{\Gamma^{2}}\partial_{t}\Delta n\ ,
\end{align}
where $\lambda_{\text{c},\alpha}=\Gamma_\alpha\left(1+f_\alpha-f_\alpha^U\right)$ is the charge relaxation rate due to coupling to lead $\alpha$, and $\lambda_\text{c}=\sum_\alpha\lambda_{\text{c},\alpha}$ is the total charge relaxation rate. While $\lambda_{\text{c},\alpha}$ are eigenvalues of the respective $W_\alpha$, the sum $\lambda_\text{c}$ is an eigenvalue of the total $W$. The dot occupation expectation value is defined as $\langle \widehat{n}\rangle=(0|\mathbf{n}|0)$.
 For arbitrary chemical potentials it reads
\begin{align}
\langle \widehat{n}\rangle & =2\left(\frac{\Gamma_{\text{L}}}{\lambda_{\text{c}}}f_\text{L}+\frac{\Gamma_{\text{R}}}{\lambda_{\text{c}}}f_\text{R}\right)\label{eq_n_dot} \ .
\end{align}
Its fluctuations are defined as $\Delta n=\langle \widehat{n}^2\rangle-\langle\widehat{n}\rangle^2=(0|\mathbf{n}^2|0)-(0|\mathbf{n}|0)^2$. For the noise expression Eq.~\eqref{eq_noise_analytic_zero_bias}, we only need the limit $\mu_\text{L}=\mu_\text{R}=\mu$, which is
\begin{align}
  \Delta n& =-k_\text{B}T\partial_\epsilon\langle\widehat{n}\rangle\ .\label{eq_Delta_n_dot}
\end{align}
These expressions yield an explicit form for the deviations from the standard stationary FRR for a single-level quantum-dot pump, which we could compactly write as~\cite{Riwar2013May}
\begin{equation}\label{eq_pumping_FR}
\begin{split}
\left.S^{(1)}_\text{LR}\right|_{\{\mu_\alpha\}\rightarrow\mu}-k_\text{B}T\left.\left(\frac{\partial I_\text{L}^{(1)}}{\partial\mu_\text{R}}+\frac{\partial I_\text{R}^{(1)}}{\partial\mu_\text{L}}\right)\right|_{\{\mu_\alpha\}\rightarrow\mu}\\ =-2k_\text{B}T\int_0^{\tau_0}\frac{dt}{\tau_0}\frac{\Gamma_\text{L}\Gamma_\text{R}}{\Gamma^2}\frac{\partial_{\epsilon}\lambda_{\text{c}}}{\lambda_\text{c}}\partial_{t}\langle\widehat{n}\rangle\ .
\end{split}
\end{equation}
We will now show that the term on the right hand side fulfills the main results presented in \ref{sec_FR_pumping}, namely that it (a)~vanishes for negligible Coulomb interaction and (b)~that it equals the nonlinear contributions to the energy-current provided by the external pumping fields.

(a)~For $U=0$ the charge relaxation rate $\lambda_\text{c}$ simply equals the coupling constant $\Gamma$. It is hence constant with respect to the dot energy, $\partial_\epsilon \lambda_\text{c}=0$, and consequently the right-hand-side of Eq.~\eqref{eq_pumping_FR} must be zero.  

\begin{figure}[t]
\centering
\includegraphics[width=0.45\textwidth]{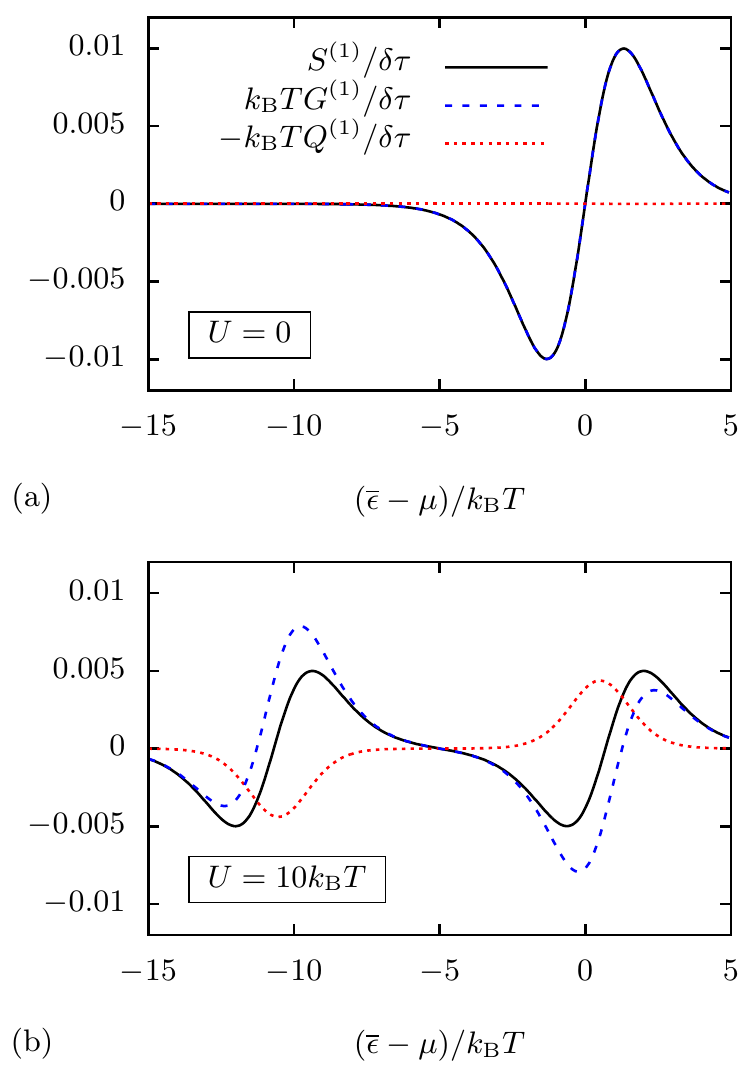}
\caption{Pumping charge current noise $S^{(1)}\equiv S^{(1)}_\text{LR}|_{\{\mu_\alpha\}\rightarrow\mu}$, conductance $G^{(1)}\equiv (\partial I_\text{L}^{(1)}/\partial\mu_\text{R}+\partial I_\text{R}^{(1)}/\partial\mu_\text{L})|_{\{\mu_\alpha\}\rightarrow\mu}$, and nonlinearity in the pumping energy input $Q^{(1)}\equiv \partial^2 I_{E,\text{pump}}^{(1)}/\partial\mu_\text{L} \partial\mu_\text{R}|_{\{\mu_\alpha\}\rightarrow\mu}$, as a function of the average energy level $\overline{\epsilon}$. The interaction strength is (a)~$U=0$ and (b) $U=10k_\text{B}T$. We choose $\epsilon(t)=\overline{\epsilon}+\delta\epsilon\sin(\Omega t+\phi)$ and $\Gamma_\text{L}=\overline{\Gamma}_\text{L}+\delta\Gamma\sin(\Omega t)$ as the driving parameters, with $\phi=\pi/2$.  The values of the parameters are $\overline{\Gamma}_\text{L}=0.65 \overline{\Gamma}$, $\Gamma_\text{R}=0.35\overline{\Gamma}$, where $\overline{\Gamma}=\overline{\Gamma}_\text{L}+\Gamma_\text{R}$. The driving parameters are $\delta\epsilon=0.1k_\text{B}T$, $\delta\Gamma=0.1\overline{\Gamma}$. The quantities are plotted with respect to the rate $1/\delta\tau\equiv\Omega\delta\Gamma_\text{L}\delta\epsilon/(\overline{\Gamma}k_\text{B}T)$.}
\label{fig_FDT_U}
\end{figure}

(b)~Based on Eq.~\eqref{eq_G_IE}, we can express the energy current provided by the external pumping fields for the single-level quantum dot as,
\begin{equation}\label{eq_IE_QD}
I_{E,\text{pump}}^{(1)}=-\int_0^{\tau_0}\frac{dt}{\tau_0}(0|\dot{\mathbf{e}}|0)=-\int_0^{\tau_0}\frac{dt}{\tau_0}\dot{\epsilon}\langle\widehat{n}\rangle\ .
\end{equation}
With straightforward algebra, one can now explicitly show that
\begin{equation}\label{eq_energy_input}
\left.\frac{\partial^2 I_{E,\text{pump}}^{(1)}}{\partial\mu_\text{L} \partial\mu_\text{R}}\right|_{\{\mu_\alpha\}\rightarrow\mu}=2\int_0^{\tau_0}\frac{dt}{\tau_0}\frac{\Gamma_\text{L}\Gamma_\text{R}}{\Gamma^2}\frac{\partial_{\epsilon}\lambda_{\text{c}}}{\lambda_{c}}\partial_{t}\langle\widehat{n}\rangle\ ,
\end{equation}
thus satisfying Eq.~\eqref{eq_FR_S1}. In Fig.~\ref{fig_FDT_U}, we plot the respective quantities $S^{(1)}\equiv S^{(1)}_\text{LR}$, $G^{(1)}\equiv\partial_{\mu_\text{R}}I_\text{L}^{(1)}+\partial_{\mu_\text{L}}I_\text{R}^{(1)}$, and $Q^{(1)}\equiv\partial_{\mu_\text{L}}\partial_{\mu_\text{R}}I_{E,\text{pump}}^{(1)}$ as a function of the time-averaged energy level $\overline{\epsilon}$. For the noninteracting system (a), we recover the equilibrium fluctuation dissipation theorem, in spite of the presence of a nonequilibrium drive. As soon as finite interactions are present (b), the correction due to the quadratic response of the pumping energy input, $I_{E,\text{pump}}^{(1)}$ is nonzero, such that the equilibrium fluctuation dissipation theorem is no longer applicable. The interaction-induced deviation from it is significant, as $I_{E,\text{pump}}^{(1)}$ is visibly of the same order of magnitude as $\partial_{\mu_\text{R}}I_\text{L}^{(1)}+\partial_{\mu_\text{L}}I_\text{R}^{(1)}$. 

\subsection{Validation of fluctuation relations through charge currents only}

Crucially, Eq.~\eqref{eq_IE_QD} indicates how the fluctuation relation in Eq.~\eqref{eq_pumping_FR} could be experimentally verified. In general, one would need the information of both the current and current noise, as well as the total energy current provided by the external pumping fields. In particular the direct measurement of energy currents may be challenging, even if current progress on fast thermometry could open up new opportunities~\cite{Gasparinetti2015Jan,Wang2018Jan}. 

However, Eq.~\eqref{eq_IE_QD} indicates that one could alternatively measure the quantum dot occupation number $\langle\widehat{n}\rangle$ as a function of time, which can be done through a local charge detector such as a quantum point contact~\cite{Fricke2013Mar,Giblin2016Jan}. 
Importantly, we note, that the time resolution need not be high. The measuring frequency $\omega$ (such that $\omega^{-1}$ can be regarded as a sampling time scale, that is, the detector resolution in time space) needs to be higher than the driving frequency $\Omega$, but in the adiabatic-response limit, can still be smaller than the tunneling rate $\Gamma_\alpha$.

The remaining quantity one needs to have access to is $\epsilon(t)$. In experiment, $\epsilon$ is typically driven through an external gate voltage. Resorting to a circuit picture, see Fig.~\ref{fig_SQD}(b), and assuming that the electrostatics of the system can be fully described by geometric capacitances, we can relate $\epsilon(t)=a\,eV_\text{g}(t)+\text{const.}$, with the lever arm factor $a=C_\text{g}/(C_\text{g}+C_\text{L}+C_\text{R})$. 
Consequently, the correction on the right-hand side of Eq.~\eqref{eq_pumping_FR} is accessible through the input function $V_\text{g}(t)$, the output function $\langle\widehat{n}\rangle(t)$ of the charge detector, and a single fitting parameter $a$.

Finally, we comment on the case where the Coulomb interaction is time-dependent, $U(t)$, and used as one of the pumping parameters~\cite{Placke2018Aug}. Within the model of geometric capacitances, where $U=e^2/(C_\text{g}+C_\text{L}+C_\text{R})$, a time-dependent $U$ can arise if the driving of the quantum dot changes the geometry, meaning that the capacitances themselves become time dependent. Also tunable effective Coulomb interaction has been realized~\cite{Hamo2016Jul}. Importantly, the general pumping fluctuation relations from Sec.~\ref{sec_symmetries_and_gauge}, see Eq.~\eqref{eq_FR_S1}, encompass also such a time-dependent driving of $U$---only the explicit shape for the pumping energy input, Eq.~(\ref{eq_energy_input}), needs to be modified. This total energy current can then be decomposed into a charge and a parity contribution, as
\begin{equation}
I_{E,\text{pump}}^{(1)}=-\int_0^{\tau_0}\frac{dt}{\tau_0}\left[\left(\dot{\epsilon}+\frac{\dot{U}}{2}\right)\langle\widehat{n}\rangle-\frac{\dot{U}}{2}\langle\widehat{p}\rangle\right]\ ,
\end{equation}
where $\widehat{p}=e^{i\pi\widehat{n}}$ is the parity operator. For this specific model, this yields the nonlinear relation $\widehat{p}=1-4\widehat{n}+2\widehat{n}^2$. It is likewise accessible by a time-resolved charge detector. Instead of measuring the total charge, one here needs to keep track of the number of single electron tunneling events, and whether they are odd or even. 
The remaining task is to experimentally control the geometry of the device to the extent that reliable information of $U(t)$ as a function of the driving parameters can be found.

\section{Conclusions}

In this work, we have studied fluctuation relations of adiabatic quantum pumps. We have shown that due to the time-dependent driving, one cannot formulate fluctuation relations of charge currents without taking into account the energy current provided by the external pumping fields. This energy current necessarily enters due to the geometric nature of the pumping observables. Namely, the elimination of the energy counting fields comes with a time-dependent unitary transformation, to which the geometric response is sensitive. 

As a result, the fluctuation response relations (FRR) of the quantum pump, obtained from derivatives of the cumulant generating function with respect to the counting fields, tie the charge current noise to the total power provided by the external driving. Interestingly, while deviations from the FRR valid for stationary systems can be expected due to pumping, we here find that they are fully induced by many-body interactions on the local quantum system.

We have concretely studied the FRR at the example of a single-level quantum dot with time-dependent energy-level and tunnel coupling. Even for this simple case the corrections to the FRR due to the pumping energy input are in general of the same order of magnitude as the pumping contributions to charge current and charge current noise, and are hence not negligible. Furthermore, we have sketched possible experimental strategies to verify our pumping fluctuation relations, even if energy currents cannot be measured. 

As an outlook, it would be interesting to extend this study to finite-time FCS and the FRR for finite-frequency observables. This could establish connections between deviations from finite-frequency FRR due to strong interactions, discovered in Ref.~\cite{Dittmann2018Sep} and first connections between finite frequency noise and energy currents established for noninteracting systems in Ref.~\cite{Moskalets2009Aug}.

\acknowledgments 
We thank F.~Haupt, M.~R.~Wegewijs and M.~Esposito for fruitful discussions. J.~S. acknowledges financial support by the Knut and Alice Wallenberg Foundation (via a Fellowship grant) and the Vetenskapsr\r{a}det, Swedish VR. R.-P.~R. acknowledges financial support by the German Federal Ministry of Education and Research within the funding program Photonic Research Germany under the contract number 13N14891.

\begin{appendix}

\section{Adiabatic expansion}\label{appendix_adiabatic}

Here, we briefly explain the adiabatic expansion for the cumulant generating function.
As a matter of fact, in spite of dealing with a dissipative open quantum system, this expansion can be envisaged in close analogy to the case of closed Hamiltonian systems~\cite{Mostafazadeh1997Mar}.

Exploiting the eigenvector decomposition of $\mathbf{W}(\{\chi_\alpha,\xi_\alpha\})$, we can expand the propagator in orders of the parameter $(k|\dot{k}')/(\lambda_k-\lambda_{k'})\simeq\Omega/\Gamma$. The lowest, zeroth order term of this FCS propagator results in
\begin{equation}
\Pi(t,t')\approx  e^{\int_{t'}^t dt_1 \left[\lambda_0(t_1)-(0|\dot{0})(t_1)\right]}|0(t))(0(t')|\ ,
\end{equation}
which is valid for sufficiently long $t-t'$, where the terms with $k>0$ will be exponentially suppressed.

Note that this zeroth order expansion of the FCS includes the first-order nonadiabatic correction of the transport equations, see Refs.~\cite{Pluecker2017Apr,Pluecker2017Nov}.

\hspace{0.0cm}
\section{Deriving FRR from fluctuation relations}\label{appendix_straightforward}

Starting from the fluctuation relations in the main text, Eqs.~\eqref{eq_FR_F0} and~\eqref{eq_FR_F1_G}, we now derive the FRR given in Eqs.~(\ref{eq_FR_I0})-(\ref{eq_FR_S1}). For this purpose, we set the counting fields $\chi_\alpha$ and $\xi_\alpha$ to zero in Eqs.~\eqref{eq_FR_F0} and~\eqref{eq_FR_F1_G}. We receive
\begin{equation}\label{eq_F0_app}
0=\mathcal{F}^{\left(0\right)}\left(\left\{ i\beta\mu_{\alpha}\right\} \right) \ ,
\end{equation}
and
\begin{equation}\label{eq_F1_app}
\beta I_{E,\text{pump}}^{\left(1\right)}=\mathcal{F}^{\left(1\right)}\left(\left\{ i\beta\mu_{\alpha}\right\} \right) \ ,
\end{equation}
where for the second equation we used the identities from Eqs.~\eqref{eq_FR_G1} and~\eqref{eq_G_IE}. As a next step, we expand both equations in a Taylor series in $\mu_\alpha$. While for $I_{E,\text{pump}}$ this expansion is trivial, note that the quantities $\mathcal{F}^{(i)}$ do not only depend on $\mu_\alpha$ through the remaining counting field arguments $\{i\beta\mu_\alpha\}$, but they also explicitly depend on $\mu_\alpha$ through the $\mu_\alpha$-dependence of the bare $W$ (i.e., in the absence of the counting fields).
Hence, the expansion up to second order in $\delta\mu_\alpha=\mu_\alpha-\mu$ provides us here with
\begin{align}
\mathcal{F}^{\left(i\right)}\left(\left\{ i\beta\mu_{\alpha}\right\} \right)&=\sum_{\alpha}\left.\delta\mu_{\alpha}\partial_{\mu_{\alpha}}\mathcal{F}\right|_{\begin{smallmatrix}\left\{ \mu_{\alpha}\right\} \rightarrow\mu\\
\left\{ \chi_{\alpha}\right\} \rightarrow0
\end{smallmatrix}}\\\nonumber
&+i\beta\sum_{\alpha}\delta\mu_{\alpha}\left.\partial_{\chi_{\alpha}}\mathcal{F}^{\left(i\right)}\right|_{\begin{smallmatrix}\left\{ \mu_{\alpha}\right\} \rightarrow\mu\\
\left\{ \chi_{\alpha}\right\} \rightarrow0
\end{smallmatrix}}\\\nonumber
&+i\beta\sum_{\alpha\gamma}\delta\mu_{\alpha}\delta\mu_{\gamma}\left.\partial_{\chi_{\alpha}}\partial_{\mu_{\gamma}}\mathcal{F}^{\left(i\right)}\right|_{\begin{smallmatrix}\left\{ \mu_{\alpha}\right\} \rightarrow\mu\\
\left\{ \chi_{\alpha}\right\} \rightarrow0
\end{smallmatrix}}\\\nonumber
&-\frac{1}{2}\beta^{2}\sum_{\alpha\gamma}\delta\mu_{\alpha}\delta\mu_{\gamma}\left.\partial_{\chi_{\alpha}}\partial_{\chi_{\gamma}}\mathcal{F}^{\left(i\right)}\right|_{\begin{smallmatrix}\left\{ \mu_{\alpha}\right\} \rightarrow\mu\\
\left\{ \chi_{\alpha}\right\} \rightarrow0
\end{smallmatrix}}\\\nonumber
&+\ldots \ .
\end{align}
Since Eqs.~\eqref{eq_F0_app} and~\eqref{eq_F1_app} have to be satisfied for arbitrary values of the chemical potentials $\mu_\alpha$, we may collect terms with the same order in $\delta\mu_\alpha$ and demand that they fulfill Eqs.~\eqref{eq_F0_app} and~\eqref{eq_F1_app} individually. Taking into account the definitions for the current and noise in Eq.~\eqref{eq_def_I} and~\eqref{eq_def_S}, we arrive at Eqs.~(\ref{eq_FR_I0})-(\ref{eq_FR_S1}).

Finally, let us point out that the relationship between the FRR in Eqs.~(\ref{eq_FR_I0}-\ref{eq_FR_S1}) and the more compact version in terms of the total pumping power, $J_\text{pump}$, see Eqs.~\eqref{eq_FRR_J1} and~\eqref{eq_FRR_J2} can be seen as follows. Based on the definition of $J_\text{pump}$ in Eq.~\eqref{eq_J_pump_def}, we see that when differentiating it with respect to $\mu_\alpha$, we get
\begin{equation}
\partial_{\mu_{\alpha}}J_{\text{pump}}=\partial_{\mu_{\alpha}}I_{E,\text{pump}}+\sum_{\gamma}\delta_{\alpha\gamma}I_{\gamma}+\sum_{\gamma}\mu_{\gamma}\partial_{\mu_{\alpha}}I_{\gamma} \ .
\end{equation}
Taking subsequently the limit of all $\mu_\alpha\rightarrow\mu$, we see that the second therm on the right hand side has to vanish due to current conservation, thus leading straight from Eq.~\eqref{eq_FRR_J1} to Eq.~\eqref{eq_FR_I1}. The very same principle applies to the second order derivative, relating Eq.~\eqref{eq_FRR_J2} to Eq.~\eqref{eq_FR_S1}, which we do not show explicitly.

\end{appendix}

\bibliography{bib_FR}

\begin{thebibliography}{75}%
\makeatletter
\providecommand \@ifxundefined [1]{%
 \@ifx{#1\undefined}
}%
\providecommand \@ifnum [1]{%
 \ifnum #1\expandafter \@firstoftwo
 \else \expandafter \@secondoftwo
 \fi
}%
\providecommand \@ifx [1]{%
 \ifx #1\expandafter \@firstoftwo
 \else \expandafter \@secondoftwo
 \fi
}%
\providecommand \natexlab [1]{#1}%
\providecommand \enquote  [1]{``#1''}%
\providecommand \bibnamefont  [1]{#1}%
\providecommand \bibfnamefont [1]{#1}%
\providecommand \citenamefont [1]{#1}%
\providecommand \href@noop [0]{\@secondoftwo}%
\providecommand \href [0]{\begingroup \@sanitize@url \@href}%
\providecommand \@href[1]{\@@startlink{#1}\@@href}%
\providecommand \@@href[1]{\endgroup#1\@@endlink}%
\providecommand \@sanitize@url [0]{\catcode `\\12\catcode `\$12\catcode
  `\&12\catcode `\#12\catcode `\^12\catcode `\_12\catcode `\%12\relax}%
\providecommand \@@startlink[1]{}%
\providecommand \@@endlink[0]{}%
\providecommand \url  [0]{\begingroup\@sanitize@url \@url }%
\providecommand \@url [1]{\endgroup\@href {#1}{\urlprefix }}%
\providecommand \urlprefix  [0]{URL }%
\providecommand \Eprint [0]{\href }%
\providecommand \doibase [0]{http://dx.doi.org/}%
\providecommand \selectlanguage [0]{\@gobble}%
\providecommand \bibinfo  [0]{\@secondoftwo}%
\providecommand \bibfield  [0]{\@secondoftwo}%
\providecommand \translation [1]{[#1]}%
\providecommand \BibitemOpen [0]{}%
\providecommand \bibitemStop [0]{}%
\providecommand \bibitemNoStop [0]{.\EOS\space}%
\providecommand \EOS [0]{\spacefactor3000\relax}%
\providecommand \BibitemShut  [1]{\csname bibitem#1\endcsname}%
\let\auto@bib@innerbib\@empty
\bibitem [{\citenamefont {Callen}\ and\ \citenamefont
  {Welton}(1951)}]{Callen1951Jul}%
  \BibitemOpen
  \bibfield  {author} {\bibinfo {author} {\bibfnamefont {H.~B.}\ \bibnamefont
  {Callen}}\ and\ \bibinfo {author} {\bibfnamefont {T.~A.}\ \bibnamefont
  {Welton}},\ }\bibfield  {title} {\enquote {\bibinfo {title} {{Irreversibility
  and Generalized Noise}},}\ }\href {\doibase 10.1103/PhysRev.83.34} {\bibfield
   {journal} {\bibinfo  {journal} {Phys. Rev.}\ }\textbf {\bibinfo {volume}
  {83}},\ \bibinfo {pages} {34--40} (\bibinfo {year} {1951})}\BibitemShut
  {NoStop}%
\bibitem [{\citenamefont {Esposito}\ and\ \citenamefont
  {Mukamel}(2006)}]{Esposito2006Apr}%
  \BibitemOpen
  \bibfield  {author} {\bibinfo {author} {\bibfnamefont {M.}~\bibnamefont
  {Esposito}}\ and\ \bibinfo {author} {\bibfnamefont {S.}~\bibnamefont
  {Mukamel}},\ }\bibfield  {title} {\enquote {\bibinfo {title} {{Fluctuation
  theorems for quantum master equations}},}\ }\href {\doibase
  10.1103/PhysRevE.73.046129} {\bibfield  {journal} {\bibinfo  {journal} {Phys.
  Rev. E}\ }\textbf {\bibinfo {volume} {73}},\ \bibinfo {pages} {046129}
  (\bibinfo {year} {2006})}\BibitemShut {NoStop}%
\bibitem [{\citenamefont {Andrieux}\ and\ \citenamefont
  {Gaspard}(2007)}]{Andrieux2007Feb}%
  \BibitemOpen
  \bibfield  {author} {\bibinfo {author} {\bibfnamefont {D.}~\bibnamefont
  {Andrieux}}\ and\ \bibinfo {author} {\bibfnamefont {P.}~\bibnamefont
  {Gaspard}},\ }\bibfield  {title} {\enquote {\bibinfo {title} {{A fluctuation
  theorem for currents and non-linear response coefficients}},}\ }\href
  {\doibase 10.1088/1742-5468/2007/02/p02006} {\bibfield  {journal} {\bibinfo
  {journal} {J. Stat. Mech.: Theory Exp.}\ }\textbf {\bibinfo {volume}
  {2007}},\ \bibinfo {pages} {P02006--P02006} (\bibinfo {year}
  {2007})}\BibitemShut {NoStop}%
\bibitem [{\citenamefont {Esposito}\ \emph {et~al.}(2007)\citenamefont
  {Esposito}, \citenamefont {Harbola},\ and\ \citenamefont
  {Mukamel}}]{Esposito2007Sep}%
  \BibitemOpen
  \bibfield  {author} {\bibinfo {author} {\bibfnamefont {M.}~\bibnamefont
  {Esposito}}, \bibinfo {author} {\bibfnamefont {U.}~\bibnamefont {Harbola}}, \
  and\ \bibinfo {author} {\bibfnamefont {S.}~\bibnamefont {Mukamel}},\
  }\bibfield  {title} {\enquote {\bibinfo {title} {{Entropy fluctuation
  theorems in driven open systems: Application to electron counting
  statistics}},}\ }\href {\doibase 10.1103/PhysRevE.76.031132} {\bibfield
  {journal} {\bibinfo  {journal} {Phys. Rev. E}\ }\textbf {\bibinfo {volume}
  {76}},\ \bibinfo {pages} {031132} (\bibinfo {year} {2007})}\BibitemShut
  {NoStop}%
\bibitem [{\citenamefont {F\"orster}\ and\ \citenamefont
  {B\"uttiker}(2008)}]{Foerster2008Sep}%
  \BibitemOpen
  \bibfield  {author} {\bibinfo {author} {\bibfnamefont {H.}~\bibnamefont
  {F\"orster}}\ and\ \bibinfo {author} {\bibfnamefont {M.}~\bibnamefont
  {B\"uttiker}},\ }\bibfield  {title} {\enquote {\bibinfo {title} {Fluctuation
  relations without microreversibility in nonlinear transport},}\ }\href
  {\doibase 10.1103/PhysRevLett.101.136805} {\bibfield  {journal} {\bibinfo
  {journal} {Phys. Rev. Lett.}\ }\textbf {\bibinfo {volume} {101}},\ \bibinfo
  {pages} {136805} (\bibinfo {year} {2008})}\BibitemShut {NoStop}%
\bibitem [{\citenamefont {Esposito}\ \emph {et~al.}(2009)\citenamefont
  {Esposito}, \citenamefont {Harbola},\ and\ \citenamefont
  {Mukamel}}]{Esposito2009Dec}%
  \BibitemOpen
  \bibfield  {author} {\bibinfo {author} {\bibfnamefont {M.}~\bibnamefont
  {Esposito}}, \bibinfo {author} {\bibfnamefont {U.}~\bibnamefont {Harbola}}, \
  and\ \bibinfo {author} {\bibfnamefont {S.}~\bibnamefont {Mukamel}},\
  }\bibfield  {title} {\enquote {\bibinfo {title} {{Nonequilibrium
  fluctuations, fluctuation theorems, and counting statistics in quantum
  systems}},}\ }\href {\doibase 10.1103/RevModPhys.81.1665} {\bibfield
  {journal} {\bibinfo  {journal} {Rev. Mod. Phys.}\ }\textbf {\bibinfo {volume}
  {81}},\ \bibinfo {pages} {1665--1702} (\bibinfo {year} {2009})}\BibitemShut
  {NoStop}%
\bibitem [{\citenamefont {Utsumi}\ \emph {et~al.}(2010)\citenamefont {Utsumi},
  \citenamefont {Golubev}, \citenamefont {Marthaler}, \citenamefont {Saito},
  \citenamefont {Fujisawa},\ and\ \citenamefont {Sch\"on}}]{Utsumi2010Mar}%
  \BibitemOpen
  \bibfield  {author} {\bibinfo {author} {\bibfnamefont {Y.}~\bibnamefont
  {Utsumi}}, \bibinfo {author} {\bibfnamefont {D.~S.}\ \bibnamefont {Golubev}},
  \bibinfo {author} {\bibfnamefont {M.}~\bibnamefont {Marthaler}}, \bibinfo
  {author} {\bibfnamefont {K.}~\bibnamefont {Saito}}, \bibinfo {author}
  {\bibfnamefont {T.}~\bibnamefont {Fujisawa}}, \ and\ \bibinfo {author}
  {\bibfnamefont {G.}~\bibnamefont {Sch\"on}},\ }\bibfield  {title} {\enquote
  {\bibinfo {title} {Bidirectional single-electron counting and the fluctuation
  theorem},}\ }\href {\doibase 10.1103/PhysRevB.81.125331} {\bibfield
  {journal} {\bibinfo  {journal} {Phys. Rev. B}\ }\textbf {\bibinfo {volume}
  {81}},\ \bibinfo {pages} {125331} (\bibinfo {year} {2010})}\BibitemShut
  {NoStop}%
\bibitem [{\citenamefont {Ren}\ \emph {et~al.}(2010)\citenamefont {Ren},
  \citenamefont {H{\ifmmode\ddot{a}\else\"{a}\fi}nggi},\ and\ \citenamefont
  {Li}}]{Ren2010Apr}%
  \BibitemOpen
  \bibfield  {author} {\bibinfo {author} {\bibfnamefont {J.}~\bibnamefont
  {Ren}}, \bibinfo {author} {\bibfnamefont {P.}~\bibnamefont
  {H{\ifmmode\ddot{a}\else\"{a}\fi}nggi}}, \ and\ \bibinfo {author}
  {\bibfnamefont {B.}~\bibnamefont {Li}},\ }\bibfield  {title} {\enquote
  {\bibinfo {title} {{Berry-Phase-Induced Heat Pumping and Its Impact on the
  Fluctuation Theorem}},}\ }\href {\doibase 10.1103/PhysRevLett.104.170601}
  {\bibfield  {journal} {\bibinfo  {journal} {Phys. Rev. Lett.}\ }\textbf
  {\bibinfo {volume} {104}},\ \bibinfo {pages} {170601} (\bibinfo {year}
  {2010})}\BibitemShut {NoStop}%
\bibitem [{\citenamefont {Campisi}\ \emph {et~al.}(2011)\citenamefont
  {Campisi}, \citenamefont {H{\ifmmode\ddot{a}\else\"{a}\fi}nggi},\ and\
  \citenamefont {Talkner}}]{Campisi2011Jul}%
  \BibitemOpen
  \bibfield  {author} {\bibinfo {author} {\bibfnamefont {M.}~\bibnamefont
  {Campisi}}, \bibinfo {author} {\bibfnamefont {P.}~\bibnamefont
  {H{\ifmmode\ddot{a}\else\"{a}\fi}nggi}}, \ and\ \bibinfo {author}
  {\bibfnamefont {P.}~\bibnamefont {Talkner}},\ }\bibfield  {title} {\enquote
  {\bibinfo {title} {{Colloquium: Quantum fluctuation relations: Foundations
  and applications}},}\ }\href {\doibase 10.1103/RevModPhys.83.771} {\bibfield
  {journal} {\bibinfo  {journal} {Rev. Mod. Phys.}\ }\textbf {\bibinfo {volume}
  {83}},\ \bibinfo {pages} {771--791} (\bibinfo {year} {2011})}\BibitemShut
  {NoStop}%
\bibitem [{\citenamefont {Saira}\ \emph {et~al.}(2012)\citenamefont {Saira},
  \citenamefont {Yoon}, \citenamefont {Tanttu}, \citenamefont
  {M{\ifmmode\ddot{o}\else\"{o}\fi}tt{\ifmmode\ddot{o}\else\"{o}\fi}nen},
  \citenamefont {Averin},\ and\ \citenamefont {Pekola}}]{Saira2012Oct}%
  \BibitemOpen
  \bibfield  {author} {\bibinfo {author} {\bibfnamefont {O.-P.}\ \bibnamefont
  {Saira}}, \bibinfo {author} {\bibfnamefont {Y.}~\bibnamefont {Yoon}},
  \bibinfo {author} {\bibfnamefont {T.}~\bibnamefont {Tanttu}}, \bibinfo
  {author} {\bibfnamefont {M.}~\bibnamefont
  {M{\ifmmode\ddot{o}\else\"{o}\fi}tt{\ifmmode\ddot{o}\else\"{o}\fi}nen}},
  \bibinfo {author} {\bibfnamefont {D.~V.}\ \bibnamefont {Averin}}, \ and\
  \bibinfo {author} {\bibfnamefont {J.~P.}\ \bibnamefont {Pekola}},\ }\bibfield
   {title} {\enquote {\bibinfo {title} {{Test of the Jarzynski and Crooks
  Fluctuation Relations in an Electronic System}},}\ }\href {\doibase
  10.1103/PhysRevLett.109.180601} {\bibfield  {journal} {\bibinfo  {journal}
  {Phys. Rev. Lett.}\ }\textbf {\bibinfo {volume} {109}},\ \bibinfo {pages}
  {180601} (\bibinfo {year} {2012})}\BibitemShut {NoStop}%
\bibitem [{\citenamefont {Lippiello}\ \emph {et~al.}(2014)\citenamefont
  {Lippiello}, \citenamefont {Baiesi},\ and\ \citenamefont
  {Sarracino}}]{Lippiello2014Apr}%
  \BibitemOpen
  \bibfield  {author} {\bibinfo {author} {\bibfnamefont {E.}~\bibnamefont
  {Lippiello}}, \bibinfo {author} {\bibfnamefont {M.}~\bibnamefont {Baiesi}}, \
  and\ \bibinfo {author} {\bibfnamefont {A.}~\bibnamefont {Sarracino}},\
  }\bibfield  {title} {\enquote {\bibinfo {title} {{Nonequilibrium
  Fluctuation-Dissipation Theorem and Heat Production}},}\ }\href {\doibase
  10.1103/PhysRevLett.112.140602} {\bibfield  {journal} {\bibinfo  {journal}
  {Phys. Rev. Lett.}\ }\textbf {\bibinfo {volume} {112}},\ \bibinfo {pages}
  {140602} (\bibinfo {year} {2014})}\BibitemShut {NoStop}%
\bibitem [{\citenamefont {Ansari}\ and\ \citenamefont
  {Nazarov}(2015)}]{Ansari2015Mar}%
  \BibitemOpen
  \bibfield  {author} {\bibinfo {author} {\bibfnamefont {M.~H.}\ \bibnamefont
  {Ansari}}\ and\ \bibinfo {author} {\bibfnamefont {Y.~V.}\ \bibnamefont
  {Nazarov}},\ }\bibfield  {title} {\enquote {\bibinfo {title}
  {{R{\ifmmode\backslash\else\textbackslash\fi}'enyi entropy flows from quantum
  heat engines}},}\ }\href {\doibase 10.1103/PhysRevB.91.104303} {\bibfield
  {journal} {\bibinfo  {journal} {Phys. Rev. B}\ }\textbf {\bibinfo {volume}
  {91}},\ \bibinfo {pages} {104303} (\bibinfo {year} {2015})}\BibitemShut
  {NoStop}%
\bibitem [{\citenamefont {Borrelli}\ \emph {et~al.}(2015)\citenamefont
  {Borrelli}, \citenamefont {Koski}, \citenamefont {Maniscalco},\ and\
  \citenamefont {Pekola}}]{Borrelli2015Jan}%
  \BibitemOpen
  \bibfield  {author} {\bibinfo {author} {\bibfnamefont {M.}~\bibnamefont
  {Borrelli}}, \bibinfo {author} {\bibfnamefont {J.~V.}\ \bibnamefont {Koski}},
  \bibinfo {author} {\bibfnamefont {S.}~\bibnamefont {Maniscalco}}, \ and\
  \bibinfo {author} {\bibfnamefont {J.~P.}\ \bibnamefont {Pekola}},\ }\bibfield
   {title} {\enquote {\bibinfo {title} {{Fluctuation relations for driven
  coupled classical two-level systems with incomplete measurements}},}\ }\href
  {\doibase 10.1103/PhysRevE.91.012145} {\bibfield  {journal} {\bibinfo
  {journal} {Phys. Rev. E}\ }\textbf {\bibinfo {volume} {91}},\ \bibinfo
  {pages} {012145} (\bibinfo {year} {2015})}\BibitemShut {NoStop}%
\bibitem [{\citenamefont {Cuetara}\ \emph {et~al.}(2015)\citenamefont
  {Cuetara}, \citenamefont {Engel},\ and\ \citenamefont
  {Esposito}}]{Cuetara2015May}%
  \BibitemOpen
  \bibfield  {author} {\bibinfo {author} {\bibfnamefont {G.~B.}\ \bibnamefont
  {Cuetara}}, \bibinfo {author} {\bibfnamefont {A.}~\bibnamefont {Engel}}, \
  and\ \bibinfo {author} {\bibfnamefont {M.}~\bibnamefont {Esposito}},\
  }\bibfield  {title} {\enquote {\bibinfo {title} {{Stochastic thermodynamics
  of rapidly driven systems}},}\ }\href {\doibase
  10.1088/1367-2630/17/5/055002} {\bibfield  {journal} {\bibinfo  {journal}
  {New J. Phys.}\ }\textbf {\bibinfo {volume} {17}},\ \bibinfo {pages} {055002}
  (\bibinfo {year} {2015})}\BibitemShut {NoStop}%
\bibitem [{\citenamefont {Altaner}\ \emph {et~al.}(2016)\citenamefont
  {Altaner}, \citenamefont {Polettini},\ and\ \citenamefont
  {Esposito}}]{Altaner2016Oct}%
  \BibitemOpen
  \bibfield  {author} {\bibinfo {author} {\bibfnamefont {B.}~\bibnamefont
  {Altaner}}, \bibinfo {author} {\bibfnamefont {M.}~\bibnamefont {Polettini}},
  \ and\ \bibinfo {author} {\bibfnamefont {M.}~\bibnamefont {Esposito}},\
  }\bibfield  {title} {\enquote {\bibinfo {title} {{Fluctuation-Dissipation
  Relations Far from Equilibrium}},}\ }\href {\doibase
  10.1103/PhysRevLett.117.180601} {\bibfield  {journal} {\bibinfo  {journal}
  {Phys. Rev. Lett.}\ }\textbf {\bibinfo {volume} {117}},\ \bibinfo {pages}
  {180601} (\bibinfo {year} {2016})}\BibitemShut {NoStop}%
\bibitem [{\citenamefont {Benito}\ \emph {et~al.}(2016)\citenamefont {Benito},
  \citenamefont {Niklas},\ and\ \citenamefont {Kohler}}]{Benito2016Nov}%
  \BibitemOpen
  \bibfield  {author} {\bibinfo {author} {\bibfnamefont {M.}~\bibnamefont
  {Benito}}, \bibinfo {author} {\bibfnamefont {M.}~\bibnamefont {Niklas}}, \
  and\ \bibinfo {author} {\bibfnamefont {S.}~\bibnamefont {Kohler}},\
  }\bibfield  {title} {\enquote {\bibinfo {title} {{Full-counting statistics of
  time-dependent conductors}},}\ }\href {\doibase 10.1103/PhysRevB.94.195433}
  {\bibfield  {journal} {\bibinfo  {journal} {Phys. Rev. B}\ }\textbf {\bibinfo
  {volume} {94}},\ \bibinfo {pages} {195433} (\bibinfo {year}
  {2016})}\BibitemShut {NoStop}%
\bibitem [{\citenamefont {Kubo}(1966)}]{Kubo1966Jan}%
  \BibitemOpen
  \bibfield  {author} {\bibinfo {author} {\bibfnamefont {R.}~\bibnamefont
  {Kubo}},\ }\bibfield  {title} {\enquote {\bibinfo {title} {{The
  fluctuation-dissipation theorem}},}\ }\href {\doibase
  10.1088/0034-4885/29/1/306} {\bibfield  {journal} {\bibinfo  {journal} {Rep.
  Prog. Phys.}\ }\textbf {\bibinfo {volume} {29}},\ \bibinfo {pages} {255--284}
  (\bibinfo {year} {1966})}\BibitemShut {NoStop}%
\bibitem [{\citenamefont {White}\ \emph {et~al.}(1996)\citenamefont {White},
  \citenamefont {Galleano}, \citenamefont {Actis}, \citenamefont {Brixy},
  \citenamefont {De~Groot}, \citenamefont {Dubbeldam}, \citenamefont {Reesink},
  \citenamefont {Edler}, \citenamefont {Sakurai}, \citenamefont {Shepard},\
  and\ \citenamefont {Gallop}}]{White1996Aug}%
  \BibitemOpen
  \bibfield  {author} {\bibinfo {author} {\bibfnamefont {D.~R.}\ \bibnamefont
  {White}}, \bibinfo {author} {\bibfnamefont {R.}~\bibnamefont {Galleano}},
  \bibinfo {author} {\bibfnamefont {A.}~\bibnamefont {Actis}}, \bibinfo
  {author} {\bibfnamefont {H.}~\bibnamefont {Brixy}}, \bibinfo {author}
  {\bibfnamefont {M.}~\bibnamefont {De~Groot}}, \bibinfo {author}
  {\bibfnamefont {J.}~\bibnamefont {Dubbeldam}}, \bibinfo {author}
  {\bibfnamefont {A.~L.}\ \bibnamefont {Reesink}}, \bibinfo {author}
  {\bibfnamefont {F.}~\bibnamefont {Edler}}, \bibinfo {author} {\bibfnamefont
  {H.}~\bibnamefont {Sakurai}}, \bibinfo {author} {\bibfnamefont {R.~L.}\
  \bibnamefont {Shepard}}, \ and\ \bibinfo {author} {\bibfnamefont {J.~C.}\
  \bibnamefont {Gallop}},\ }\bibfield  {title} {\enquote {\bibinfo {title}
  {{The status of Johnson noise thermometry}},}\ }\href {\doibase
  10.1088/0026-1394/33/4/6} {\bibfield  {journal} {\bibinfo  {journal}
  {Metrologia}\ }\textbf {\bibinfo {volume} {33}},\ \bibinfo {pages} {325--335}
  (\bibinfo {year} {1996})}\BibitemShut {NoStop}%
\bibitem [{\citenamefont {Ren}\ and\ \citenamefont
  {Sinitsyn}(2013)}]{Ren2013May}%
  \BibitemOpen
  \bibfield  {author} {\bibinfo {author} {\bibfnamefont {J.}~\bibnamefont
  {Ren}}\ and\ \bibinfo {author} {\bibfnamefont {N.~A.}\ \bibnamefont
  {Sinitsyn}},\ }\bibfield  {title} {\enquote {\bibinfo {title} {{Braid group
  and topological phase transitions in nonequilibrium stochastic dynamics}},}\
  }\href {\doibase 10.1103/PhysRevE.87.050101} {\bibfield  {journal} {\bibinfo
  {journal} {Phys. Rev. E}\ }\textbf {\bibinfo {volume} {87}},\ \bibinfo
  {pages} {050101} (\bibinfo {year} {2013})}\BibitemShut {NoStop}%
\bibitem [{\citenamefont {Riwar}(2019)}]{Riwar2019Dec}%
  \BibitemOpen
  \bibfield  {author} {\bibinfo {author} {\bibfnamefont {R.-P.}\ \bibnamefont
  {Riwar}},\ }\bibfield  {title} {\enquote {\bibinfo {title} {{Fractional
  charges in conventional sequential electron tunneling}},}\ }\href {\doibase
  10.1103/PhysRevB.100.245416} {\bibfield  {journal} {\bibinfo  {journal}
  {Phys. Rev. B}\ }\textbf {\bibinfo {volume} {100}},\ \bibinfo {pages}
  {245416} (\bibinfo {year} {2019})}\BibitemShut {NoStop}%
\bibitem [{\citenamefont {Moskalets}\ and\ \citenamefont
  {B{\ifmmode\ddot{u}\else\"{u}\fi}ttiker}(2009)}]{Moskalets2009Aug}%
  \BibitemOpen
  \bibfield  {author} {\bibinfo {author} {\bibfnamefont {M.}~\bibnamefont
  {Moskalets}}\ and\ \bibinfo {author} {\bibfnamefont {M.}~\bibnamefont
  {B{\ifmmode\ddot{u}\else\"{u}\fi}ttiker}},\ }\bibfield  {title} {\enquote
  {\bibinfo {title} {{Heat production and current noise for single- and
  double-cavity quantum capacitors}},}\ }\href {\doibase
  10.1103/PhysRevB.80.081302} {\bibfield  {journal} {\bibinfo  {journal} {Phys.
  Rev. B}\ }\textbf {\bibinfo {volume} {80}},\ \bibinfo {pages} {081302}
  (\bibinfo {year} {2009})}\BibitemShut {NoStop}%
\bibitem [{\citenamefont {Bulnes~Cuetara}\ \emph {et~al.}(2014)\citenamefont
  {Bulnes~Cuetara}, \citenamefont {Esposito},\ and\ \citenamefont
  {Imparato}}]{Bulnes2014May}%
  \BibitemOpen
  \bibfield  {author} {\bibinfo {author} {\bibfnamefont {G.}~\bibnamefont
  {Bulnes~Cuetara}}, \bibinfo {author} {\bibfnamefont {M.}~\bibnamefont
  {Esposito}}, \ and\ \bibinfo {author} {\bibfnamefont {A.}~\bibnamefont
  {Imparato}},\ }\bibfield  {title} {\enquote {\bibinfo {title} {Exact
  fluctuation theorem without ensemble quantities},}\ }\href {\doibase
  10.1103/PhysRevE.89.052119} {\bibfield  {journal} {\bibinfo  {journal} {Phys.
  Rev. E}\ }\textbf {\bibinfo {volume} {89}},\ \bibinfo {pages} {052119}
  (\bibinfo {year} {2014})}\BibitemShut {NoStop}%
\bibitem [{\citenamefont {Croy}\ and\ \citenamefont
  {Saalmann}(2016)}]{Croy2016Apr}%
  \BibitemOpen
  \bibfield  {author} {\bibinfo {author} {\bibfnamefont {A.}~\bibnamefont
  {Croy}}\ and\ \bibinfo {author} {\bibfnamefont {U.}~\bibnamefont
  {Saalmann}},\ }\bibfield  {title} {\enquote {\bibinfo {title} {{Full counting
  statistics of a nonadiabatic electron pump}},}\ }\href {\doibase
  10.1103/PhysRevB.93.165428} {\bibfield  {journal} {\bibinfo  {journal} {Phys.
  Rev. B}\ }\textbf {\bibinfo {volume} {93}},\ \bibinfo {pages} {165428}
  (\bibinfo {year} {2016})}\BibitemShut {NoStop}%
\bibitem [{\citenamefont {Watanabe}\ and\ \citenamefont
  {Hayakawa}(2017)}]{Watanabe2017Aug}%
  \BibitemOpen
  \bibfield  {author} {\bibinfo {author} {\bibfnamefont {K.~L.}\ \bibnamefont
  {Watanabe}}\ and\ \bibinfo {author} {\bibfnamefont {H.}~\bibnamefont
  {Hayakawa}},\ }\bibfield  {title} {\enquote {\bibinfo {title} {{Geometric
  fluctuation theorem for a spin-boson system}},}\ }\href {\doibase
  10.1103/PhysRevE.96.022118} {\bibfield  {journal} {\bibinfo  {journal} {Phys.
  Rev. E}\ }\textbf {\bibinfo {volume} {96}},\ \bibinfo {pages} {022118}
  (\bibinfo {year} {2017})}\BibitemShut {NoStop}%
\bibitem [{\citenamefont {Hino}\ and\ \citenamefont
  {Hayakawa}(2020)}]{Hino2020Jul}%
  \BibitemOpen
  \bibfield  {author} {\bibinfo {author} {\bibfnamefont {Y.}~\bibnamefont
  {Hino}}\ and\ \bibinfo {author} {\bibfnamefont {H.}~\bibnamefont
  {Hayakawa}},\ }\bibfield  {title} {\enquote {\bibinfo {title} {{Fluctuation
  relations for adiabatic pumping}},}\ }\href {\doibase
  10.1103/PhysRevE.102.012115} {\bibfield  {journal} {\bibinfo  {journal}
  {Phys. Rev. E}\ }\textbf {\bibinfo {volume} {102}},\ \bibinfo {pages}
  {012115} (\bibinfo {year} {2020})}\BibitemShut {NoStop}%
\bibitem [{\citenamefont {Flindt}\ \emph {et~al.}(2009)\citenamefont {Flindt},
  \citenamefont {Fricke}, \citenamefont {Hohls}, \citenamefont
  {Novotn{\ifmmode\acute{y}\else\'{y}\fi}}, \citenamefont
  {Neto{\ifmmode\check{c}\else\v{c}\fi}n{\ifmmode\acute{y}\else\'{y}\fi}},
  \citenamefont {Brandes},\ and\ \citenamefont {Haug}}]{Flindt2009Jun}%
  \BibitemOpen
  \bibfield  {author} {\bibinfo {author} {\bibfnamefont {C.}~\bibnamefont
  {Flindt}}, \bibinfo {author} {\bibfnamefont {C.}~\bibnamefont {Fricke}},
  \bibinfo {author} {\bibfnamefont {F.}~\bibnamefont {Hohls}}, \bibinfo
  {author} {\bibfnamefont {T.}~\bibnamefont
  {Novotn{\ifmmode\acute{y}\else\'{y}\fi}}}, \bibinfo {author} {\bibfnamefont
  {K.}~\bibnamefont
  {Neto{\ifmmode\check{c}\else\v{c}\fi}n{\ifmmode\acute{y}\else\'{y}\fi}}},
  \bibinfo {author} {\bibfnamefont {T.}~\bibnamefont {Brandes}}, \ and\
  \bibinfo {author} {\bibfnamefont {R.~J.}\ \bibnamefont {Haug}},\ }\bibfield
  {title} {\enquote {\bibinfo {title} {{Universal oscillations in counting
  statistics}},}\ }\href {\doibase 10.1073/pnas.0901002106} {\bibfield
  {journal} {\bibinfo  {journal} {Proc. Natl. Acad. Sci. U.S.A.}\ }\textbf
  {\bibinfo {volume} {106}},\ \bibinfo {pages} {10116--10119} (\bibinfo {year}
  {2009})}\BibitemShut {NoStop}%
\bibitem [{\citenamefont {Parmentier}\ \emph {et~al.}(2012)\citenamefont
  {Parmentier}, \citenamefont {Bocquillon}, \citenamefont {Berroir},
  \citenamefont {Glattli}, \citenamefont
  {Pla{\ifmmode\mbox{\c{c}}\else\c{c}\fi}ais}, \citenamefont
  {F{\ifmmode\grave{e}\else\`{e}\fi}ve}, \citenamefont {Albert}, \citenamefont
  {Flindt},\ and\ \citenamefont
  {B{\ifmmode\ddot{u}\else\"{u}\fi}ttiker}}]{Parmentier2012Apr}%
  \BibitemOpen
  \bibfield  {author} {\bibinfo {author} {\bibfnamefont {F.~D.}\ \bibnamefont
  {Parmentier}}, \bibinfo {author} {\bibfnamefont {E.}~\bibnamefont
  {Bocquillon}}, \bibinfo {author} {\bibfnamefont {J.-M.}\ \bibnamefont
  {Berroir}}, \bibinfo {author} {\bibfnamefont {D.~C.}\ \bibnamefont
  {Glattli}}, \bibinfo {author} {\bibfnamefont {B.}~\bibnamefont
  {Pla{\ifmmode\mbox{\c{c}}\else\c{c}\fi}ais}}, \bibinfo {author}
  {\bibfnamefont {G.}~\bibnamefont {F{\ifmmode\grave{e}\else\`{e}\fi}ve}},
  \bibinfo {author} {\bibfnamefont {M.}~\bibnamefont {Albert}}, \bibinfo
  {author} {\bibfnamefont {C.}~\bibnamefont {Flindt}}, \ and\ \bibinfo {author}
  {\bibfnamefont {M.}~\bibnamefont {B{\ifmmode\ddot{u}\else\"{u}\fi}ttiker}},\
  }\bibfield  {title} {\enquote {\bibinfo {title} {{Current noise spectrum of a
  single-particle emitter: Theory and experiment}},}\ }\href {\doibase
  10.1103/PhysRevB.85.165438} {\bibfield  {journal} {\bibinfo  {journal} {Phys.
  Rev. B}\ }\textbf {\bibinfo {volume} {85}},\ \bibinfo {pages} {165438}
  (\bibinfo {year} {2012})}\BibitemShut {NoStop}%
\bibitem [{\citenamefont {Bocquillon}\ \emph {et~al.}(2012)\citenamefont
  {Bocquillon}, \citenamefont {Parmentier}, \citenamefont {Grenier},
  \citenamefont {Berroir}, \citenamefont {Degiovanni}, \citenamefont {Glattli},
  \citenamefont {Pla\ifmmode~\mbox{\c{c}}\else \c{c}\fi{}ais}, \citenamefont
  {Cavanna}, \citenamefont {Jin},\ and\ \citenamefont
  {F\`eve}}]{Bocquillon2012May}%
  \BibitemOpen
  \bibfield  {author} {\bibinfo {author} {\bibfnamefont {E.}~\bibnamefont
  {Bocquillon}}, \bibinfo {author} {\bibfnamefont {F.~D.}\ \bibnamefont
  {Parmentier}}, \bibinfo {author} {\bibfnamefont {C.}~\bibnamefont {Grenier}},
  \bibinfo {author} {\bibfnamefont {J.-M.}\ \bibnamefont {Berroir}}, \bibinfo
  {author} {\bibfnamefont {P.}~\bibnamefont {Degiovanni}}, \bibinfo {author}
  {\bibfnamefont {D.~C.}\ \bibnamefont {Glattli}}, \bibinfo {author}
  {\bibfnamefont {B.}~\bibnamefont {Pla\ifmmode~\mbox{\c{c}}\else
  \c{c}\fi{}ais}}, \bibinfo {author} {\bibfnamefont {A.}~\bibnamefont
  {Cavanna}}, \bibinfo {author} {\bibfnamefont {Y.}~\bibnamefont {Jin}}, \ and\
  \bibinfo {author} {\bibfnamefont {G.}~\bibnamefont {F\`eve}},\ }\bibfield
  {title} {\enquote {\bibinfo {title} {Electron quantum optics: Partitioning
  electrons one by one},}\ }\href {\doibase 10.1103/PhysRevLett.108.196803}
  {\bibfield  {journal} {\bibinfo  {journal} {Phys. Rev. Lett.}\ }\textbf
  {\bibinfo {volume} {108}},\ \bibinfo {pages} {196803} (\bibinfo {year}
  {2012})}\BibitemShut {NoStop}%
\bibitem [{\citenamefont {Bocquillon}\ \emph {et~al.}(2013)\citenamefont
  {Bocquillon}, \citenamefont {Freulon}, \citenamefont {Berroir}, \citenamefont
  {Degiovanni}, \citenamefont {Pla{\ifmmode\mbox{\c{c}}\else\c{c}\fi}ais},
  \citenamefont {Cavanna}, \citenamefont {Jin},\ and\ \citenamefont
  {F{\ifmmode\grave{e}\else\`{e}\fi}ve}}]{Bocquillon2013Mar}%
  \BibitemOpen
  \bibfield  {author} {\bibinfo {author} {\bibfnamefont {E.}~\bibnamefont
  {Bocquillon}}, \bibinfo {author} {\bibfnamefont {V.}~\bibnamefont {Freulon}},
  \bibinfo {author} {\bibfnamefont {J.-M.}\ \bibnamefont {Berroir}}, \bibinfo
  {author} {\bibfnamefont {P.}~\bibnamefont {Degiovanni}}, \bibinfo {author}
  {\bibfnamefont {B.}~\bibnamefont
  {Pla{\ifmmode\mbox{\c{c}}\else\c{c}\fi}ais}}, \bibinfo {author}
  {\bibfnamefont {A.}~\bibnamefont {Cavanna}}, \bibinfo {author} {\bibfnamefont
  {Y.}~\bibnamefont {Jin}}, \ and\ \bibinfo {author} {\bibfnamefont
  {G.}~\bibnamefont {F{\ifmmode\grave{e}\else\`{e}\fi}ve}},\ }\bibfield
  {title} {\enquote {\bibinfo {title} {{Coherence and Indistinguishability of
  Single Electrons Emitted by Independent Sources}},}\ }\href {\doibase
  10.1126/science.1232572} {\bibfield  {journal} {\bibinfo  {journal}
  {Science}\ }\textbf {\bibinfo {volume} {339}},\ \bibinfo {pages} {1054--1057}
  (\bibinfo {year} {2013})}\BibitemShut {NoStop}%
\bibitem [{\citenamefont {Gabelli}\ and\ \citenamefont
  {Reulet}(2013)}]{Gabelli2013Feb}%
  \BibitemOpen
  \bibfield  {author} {\bibinfo {author} {\bibfnamefont {J.}~\bibnamefont
  {Gabelli}}\ and\ \bibinfo {author} {\bibfnamefont {B.}~\bibnamefont
  {Reulet}},\ }\bibfield  {title} {\enquote {\bibinfo {title} {{Shaping a
  time-dependent excitation to minimize the shot noise in a tunnel
  junction}},}\ }\href {\doibase 10.1103/PhysRevB.87.075403} {\bibfield
  {journal} {\bibinfo  {journal} {Phys. Rev. B}\ }\textbf {\bibinfo {volume}
  {87}},\ \bibinfo {pages} {075403} (\bibinfo {year} {2013})}\BibitemShut
  {NoStop}%
\bibitem [{\citenamefont {Dubois}\ \emph {et~al.}(2013)\citenamefont {Dubois},
  \citenamefont {Jullien}, \citenamefont {Portier}, \citenamefont {Roche},
  \citenamefont {Cavanna}, \citenamefont {Jin}, \citenamefont {Wegscheider},
  \citenamefont {Roulleau},\ and\ \citenamefont {Glattli}}]{Dubois2013Oct}%
  \BibitemOpen
  \bibfield  {author} {\bibinfo {author} {\bibfnamefont {J.}~\bibnamefont
  {Dubois}}, \bibinfo {author} {\bibfnamefont {T.}~\bibnamefont {Jullien}},
  \bibinfo {author} {\bibfnamefont {F.}~\bibnamefont {Portier}}, \bibinfo
  {author} {\bibfnamefont {P.}~\bibnamefont {Roche}}, \bibinfo {author}
  {\bibfnamefont {A.}~\bibnamefont {Cavanna}}, \bibinfo {author} {\bibfnamefont
  {Y.}~\bibnamefont {Jin}}, \bibinfo {author} {\bibfnamefont {W.}~\bibnamefont
  {Wegscheider}}, \bibinfo {author} {\bibfnamefont {P.}~\bibnamefont
  {Roulleau}}, \ and\ \bibinfo {author} {\bibfnamefont {D.~C.}\ \bibnamefont
  {Glattli}},\ }\bibfield  {title} {\enquote {\bibinfo {title}
  {{Minimal-excitation states for electron quantum optics using levitons}},}\
  }\href {\doibase 10.1038/nature12713} {\bibfield  {journal} {\bibinfo
  {journal} {Nature}\ }\textbf {\bibinfo {volume} {502}},\ \bibinfo {pages}
  {659--663} (\bibinfo {year} {2013})}\BibitemShut {NoStop}%
\bibitem [{\citenamefont {Jullien}\ \emph {et~al.}(2014)\citenamefont
  {Jullien}, \citenamefont {Roulleau}, \citenamefont {Roche}, \citenamefont
  {Cavanna}, \citenamefont {Jin},\ and\ \citenamefont
  {Glattli}}]{Jullien2014Oct}%
  \BibitemOpen
  \bibfield  {author} {\bibinfo {author} {\bibfnamefont {T.}~\bibnamefont
  {Jullien}}, \bibinfo {author} {\bibfnamefont {P.}~\bibnamefont {Roulleau}},
  \bibinfo {author} {\bibfnamefont {B.}~\bibnamefont {Roche}}, \bibinfo
  {author} {\bibfnamefont {A.}~\bibnamefont {Cavanna}}, \bibinfo {author}
  {\bibfnamefont {Y.}~\bibnamefont {Jin}}, \ and\ \bibinfo {author}
  {\bibfnamefont {D.~C.}\ \bibnamefont {Glattli}},\ }\bibfield  {title}
  {\enquote {\bibinfo {title} {{Quantum tomography of an electron}},}\ }\href
  {\doibase 10.1038/nature13821} {\bibfield  {journal} {\bibinfo  {journal}
  {Nature}\ }\textbf {\bibinfo {volume} {514}},\ \bibinfo {pages} {603--607}
  (\bibinfo {year} {2014})}\BibitemShut {NoStop}%
\bibitem [{\citenamefont {Fricke}\ \emph {et~al.}(2013)\citenamefont {Fricke},
  \citenamefont {Wulf}, \citenamefont {Kaestner}, \citenamefont {Kashcheyevs},
  \citenamefont {Timoshenko}, \citenamefont {Nazarov}, \citenamefont {Hohls},
  \citenamefont {Mirovsky}, \citenamefont {Mackrodt}, \citenamefont {Dolata},
  \citenamefont {Weimann}, \citenamefont {Pierz},\ and\ \citenamefont
  {Schumacher}}]{Fricke2013Mar}%
  \BibitemOpen
  \bibfield  {author} {\bibinfo {author} {\bibfnamefont {L.}~\bibnamefont
  {Fricke}}, \bibinfo {author} {\bibfnamefont {M.}~\bibnamefont {Wulf}},
  \bibinfo {author} {\bibfnamefont {B.}~\bibnamefont {Kaestner}}, \bibinfo
  {author} {\bibfnamefont {V.}~\bibnamefont {Kashcheyevs}}, \bibinfo {author}
  {\bibfnamefont {J.}~\bibnamefont {Timoshenko}}, \bibinfo {author}
  {\bibfnamefont {P.}~\bibnamefont {Nazarov}}, \bibinfo {author} {\bibfnamefont
  {F.}~\bibnamefont {Hohls}}, \bibinfo {author} {\bibfnamefont
  {P.}~\bibnamefont {Mirovsky}}, \bibinfo {author} {\bibfnamefont
  {B.}~\bibnamefont {Mackrodt}}, \bibinfo {author} {\bibfnamefont
  {R.}~\bibnamefont {Dolata}}, \bibinfo {author} {\bibfnamefont
  {T.}~\bibnamefont {Weimann}}, \bibinfo {author} {\bibfnamefont
  {K.}~\bibnamefont {Pierz}}, \ and\ \bibinfo {author} {\bibfnamefont {H.~W.}\
  \bibnamefont {Schumacher}},\ }\bibfield  {title} {\enquote {\bibinfo {title}
  {{Counting Statistics for Electron Capture in a Dynamic Quantum Dot}},}\
  }\href {\doibase 10.1103/PhysRevLett.110.126803} {\bibfield  {journal}
  {\bibinfo  {journal} {Phys. Rev. Lett.}\ }\textbf {\bibinfo {volume} {110}},\
  \bibinfo {pages} {126803} (\bibinfo {year} {2013})}\BibitemShut {NoStop}%
\bibitem [{\citenamefont {Giblin}\ \emph {et~al.}(2016)\citenamefont {Giblin},
  \citenamefont {See}, \citenamefont {Petrie}, \citenamefont {Janssen},
  \citenamefont {Farrer}, \citenamefont {Griffiths}, \citenamefont {Jones},
  \citenamefont {Ritchie},\ and\ \citenamefont {Kataoka}}]{Giblin2016Jan}%
  \BibitemOpen
  \bibfield  {author} {\bibinfo {author} {\bibfnamefont {S.~P.}\ \bibnamefont
  {Giblin}}, \bibinfo {author} {\bibfnamefont {P.}~\bibnamefont {See}},
  \bibinfo {author} {\bibfnamefont {A.}~\bibnamefont {Petrie}}, \bibinfo
  {author} {\bibfnamefont {T.~J. B.~M.}\ \bibnamefont {Janssen}}, \bibinfo
  {author} {\bibfnamefont {I.}~\bibnamefont {Farrer}}, \bibinfo {author}
  {\bibfnamefont {J.~P.}\ \bibnamefont {Griffiths}}, \bibinfo {author}
  {\bibfnamefont {G.~A.~C.}\ \bibnamefont {Jones}}, \bibinfo {author}
  {\bibfnamefont {D.~A.}\ \bibnamefont {Ritchie}}, \ and\ \bibinfo {author}
  {\bibfnamefont {M.}~\bibnamefont {Kataoka}},\ }\bibfield  {title} {\enquote
  {\bibinfo {title} {{High-resolution error detection in the capture process of
  a single-electron pump}},}\ }\href {\doibase 10.1063/1.4939250} {\bibfield
  {journal} {\bibinfo  {journal} {Appl. Phys. Lett.}\ }\textbf {\bibinfo
  {volume} {108}},\ \bibinfo {pages} {023502} (\bibinfo {year}
  {2016})}\BibitemShut {NoStop}%
\bibitem [{\citenamefont {Riwar}\ \emph {et~al.}(2013)\citenamefont {Riwar},
  \citenamefont {Splettstoesser},\ and\ \citenamefont
  {K\"onig}}]{Riwar2013May}%
  \BibitemOpen
  \bibfield  {author} {\bibinfo {author} {\bibfnamefont {R.-P.}\ \bibnamefont
  {Riwar}}, \bibinfo {author} {\bibfnamefont {J.}~\bibnamefont
  {Splettstoesser}}, \ and\ \bibinfo {author} {\bibfnamefont {J.}~\bibnamefont
  {K\"onig}},\ }\bibfield  {title} {\enquote {\bibinfo {title} {Zero-frequency
  noise in adiabatically driven interacting quantum systems},}\ }\href
  {\doibase 10.1103/PhysRevB.87.195407} {\bibfield  {journal} {\bibinfo
  {journal} {Phys. Rev. B}\ }\textbf {\bibinfo {volume} {87}},\ \bibinfo
  {pages} {195407} (\bibinfo {year} {2013})}\BibitemShut {NoStop}%
\bibitem [{\citenamefont {Dittmann}\ and\ \citenamefont
  {Splettstoesser}(2018)}]{Dittmann2018Sep}%
  \BibitemOpen
  \bibfield  {author} {\bibinfo {author} {\bibfnamefont {N.}~\bibnamefont
  {Dittmann}}\ and\ \bibinfo {author} {\bibfnamefont {J.}~\bibnamefont
  {Splettstoesser}},\ }\bibfield  {title} {\enquote {\bibinfo {title}
  {{Finite-frequency noise of interacting single-electron emitters:
  Spectroscopy with higher noise harmonics}},}\ }\href {\doibase
  10.1103/PhysRevB.98.115414} {\bibfield  {journal} {\bibinfo  {journal} {Phys.
  Rev. B}\ }\textbf {\bibinfo {volume} {98}},\ \bibinfo {pages} {115414}
  (\bibinfo {year} {2018})}\BibitemShut {NoStop}%
\bibitem [{\citenamefont {Berry}(1984)}]{Berry1984Mar}%
  \BibitemOpen
  \bibfield  {author} {\bibinfo {author} {\bibfnamefont {M.~V.}\ \bibnamefont
  {Berry}},\ }\bibfield  {title} {\enquote {\bibinfo {title} {Quantal phase
  factors accompanying adiabatic changes},}\ }\href {\doibase
  10.1098/rspa.1984.0023} {\bibfield  {journal} {\bibinfo  {journal}
  {Proceedings of the Royal Society of London. A. Mathematical and Physical
  Sciences}\ }\textbf {\bibinfo {volume} {392}},\ \bibinfo {pages} {45--57}
  (\bibinfo {year} {1984})}\BibitemShut {NoStop}%
\bibitem [{\citenamefont {Uhlmann}(1986)}]{Uhlmann1986Oct}%
  \BibitemOpen
  \bibfield  {author} {\bibinfo {author} {\bibfnamefont {A.}~\bibnamefont
  {Uhlmann}},\ }\bibfield  {title} {\enquote {\bibinfo {title} {{Parallel
  transport and {\textquotedblleft}quantum holonomy{\textquotedblright} along
  density operators}},}\ }\href {\doibase 10.1016/0034-4877(86)90055-8}
  {\bibfield  {journal} {\bibinfo  {journal} {Rep. Math. Phys.}\ }\textbf
  {\bibinfo {volume} {24}},\ \bibinfo {pages} {229--240} (\bibinfo {year}
  {1986})}\BibitemShut {NoStop}%
\bibitem [{\citenamefont {Ning}\ and\ \citenamefont
  {Haken}(1992)}]{Ning1992Oct}%
  \BibitemOpen
  \bibfield  {author} {\bibinfo {author} {\bibfnamefont {C.~Z.}\ \bibnamefont
  {Ning}}\ and\ \bibinfo {author} {\bibfnamefont {H.}~\bibnamefont {Haken}},\
  }\bibfield  {title} {\enquote {\bibinfo {title} {{The geometric phase in
  nonlinear dissipative systems}},}\ }\href {\doibase
  10.1142/S0217984992001265} {\bibfield  {journal} {\bibinfo  {journal} {Mod.
  Phys. Lett. B}\ }\textbf {\bibinfo {volume} {06}},\ \bibinfo {pages}
  {1541--1568} (\bibinfo {year} {1992})}\BibitemShut {NoStop}%
\bibitem [{\citenamefont {Landsberg}(1992)}]{Landsberg1992Aug}%
  \BibitemOpen
  \bibfield  {author} {\bibinfo {author} {\bibfnamefont {A.~S.}\ \bibnamefont
  {Landsberg}},\ }\bibfield  {title} {\enquote {\bibinfo {title} {{Geometrical
  phases and symmetries in dissipative systems}},}\ }\href {\doibase
  10.1103/PhysRevLett.69.865} {\bibfield  {journal} {\bibinfo  {journal} {Phys.
  Rev. Lett.}\ }\textbf {\bibinfo {volume} {69}},\ \bibinfo {pages} {865--868}
  (\bibinfo {year} {1992})}\BibitemShut {NoStop}%
\bibitem [{\citenamefont {Brouwer}(1998)}]{Brouwer1998Oct}%
  \BibitemOpen
  \bibfield  {author} {\bibinfo {author} {\bibfnamefont {P.~W.}\ \bibnamefont
  {Brouwer}},\ }\bibfield  {title} {\enquote {\bibinfo {title} {{Scattering
  approach to parametric pumping}},}\ }\href {\doibase
  10.1103/PhysRevB.58.R10135} {\bibfield  {journal} {\bibinfo  {journal} {Phys.
  Rev. B}\ }\textbf {\bibinfo {volume} {58}},\ \bibinfo {pages}
  {R10135--R10138(R)} (\bibinfo {year} {1998})}\BibitemShut {NoStop}%
\bibitem [{\citenamefont {Avron}\ \emph {et~al.}(2001)\citenamefont {Avron},
  \citenamefont {Elgart}, \citenamefont {Graf},\ and\ \citenamefont
  {Sadun}}]{Avron2001Nov}%
  \BibitemOpen
  \bibfield  {author} {\bibinfo {author} {\bibfnamefont {J.~E.}\ \bibnamefont
  {Avron}}, \bibinfo {author} {\bibfnamefont {A.}~\bibnamefont {Elgart}},
  \bibinfo {author} {\bibfnamefont {G.~M.}\ \bibnamefont {Graf}}, \ and\
  \bibinfo {author} {\bibfnamefont {L.}~\bibnamefont {Sadun}},\ }\bibfield
  {title} {\enquote {\bibinfo {title} {Optimal quantum pumps},}\ }\href
  {\doibase 10.1103/PhysRevLett.87.236601} {\bibfield  {journal} {\bibinfo
  {journal} {Phys. Rev. Lett.}\ }\textbf {\bibinfo {volume} {87}},\ \bibinfo
  {pages} {236601} (\bibinfo {year} {2001})}\BibitemShut {NoStop}%
\bibitem [{\citenamefont {Sarandy}\ and\ \citenamefont
  {Lidar}(2005)}]{Sarandy2005Jan}%
  \BibitemOpen
  \bibfield  {author} {\bibinfo {author} {\bibfnamefont {M.~S.}\ \bibnamefont
  {Sarandy}}\ and\ \bibinfo {author} {\bibfnamefont {D.~A.}\ \bibnamefont
  {Lidar}},\ }\bibfield  {title} {\enquote {\bibinfo {title} {{Adiabatic
  approximation in open quantum systems}},}\ }\href {\doibase
  10.1103/PhysRevA.71.012331} {\bibfield  {journal} {\bibinfo  {journal} {Phys.
  Rev. A}\ }\textbf {\bibinfo {volume} {71}},\ \bibinfo {pages} {012331}
  (\bibinfo {year} {2005})}\BibitemShut {NoStop}%
\bibitem [{\citenamefont {Sinitsyn}\ and\ \citenamefont
  {Nemenman}(2007{\natexlab{a}})}]{Sinitsyn2007Nov}%
  \BibitemOpen
  \bibfield  {author} {\bibinfo {author} {\bibfnamefont {N.~A.}\ \bibnamefont
  {Sinitsyn}}\ and\ \bibinfo {author} {\bibfnamefont {I.}~\bibnamefont
  {Nemenman}},\ }\bibfield  {title} {\enquote {\bibinfo {title} {{Universal
  Geometric Theory of Mesoscopic Stochastic Pumps and Reversible Ratchets}},}\
  }\href {\doibase 10.1103/PhysRevLett.99.220408} {\bibfield  {journal}
  {\bibinfo  {journal} {Phys. Rev. Lett.}\ }\textbf {\bibinfo {volume} {99}},\
  \bibinfo {pages} {220408} (\bibinfo {year} {2007}{\natexlab{a}})}\BibitemShut
  {NoStop}%
\bibitem [{\citenamefont {Sinitsyn}\ and\ \citenamefont
  {Nemenman}(2007{\natexlab{b}})}]{Sinitsyn2007Feb}%
  \BibitemOpen
  \bibfield  {author} {\bibinfo {author} {\bibfnamefont {N.~A.}\ \bibnamefont
  {Sinitsyn}}\ and\ \bibinfo {author} {\bibfnamefont {I.}~\bibnamefont
  {Nemenman}},\ }\bibfield  {title} {\enquote {\bibinfo {title} {{The Berry
  phase and the pump flux in stochastic chemical kinetics}},}\ }\href {\doibase
  10.1209/0295-5075/77/58001} {\bibfield  {journal} {\bibinfo  {journal} {EPL}\
  }\textbf {\bibinfo {volume} {77}},\ \bibinfo {pages} {58001} (\bibinfo {year}
  {2007}{\natexlab{b}})}\BibitemShut {NoStop}%
\bibitem [{\citenamefont {Yoshii}\ and\ \citenamefont
  {Hayakawa}(2013)}]{Yoshii2013Dec}%
  \BibitemOpen
  \bibfield  {author} {\bibinfo {author} {\bibfnamefont {R.}~\bibnamefont
  {Yoshii}}\ and\ \bibinfo {author} {\bibfnamefont {H.}~\bibnamefont
  {Hayakawa}},\ }\bibfield  {title} {\enquote {\bibinfo {title} {{Analytical
  expression of geometrical pumping for a quantum dot based on quantum master
  equation}},}\ }\href {https://arxiv.org/abs/1312.3772v1} {\bibfield
  {journal} {\bibinfo  {journal} {arXiv}\ } (\bibinfo {year} {2013})},\ \Eprint
  {http://arxiv.org/abs/1312.3772} {1312.3772} \BibitemShut {NoStop}%
\bibitem [{\citenamefont {Pluecker}\ \emph
  {et~al.}(2017{\natexlab{a}})\citenamefont {Pluecker}, \citenamefont
  {Wegewijs},\ and\ \citenamefont {Splettstoesser}}]{Pluecker2017Apr}%
  \BibitemOpen
  \bibfield  {author} {\bibinfo {author} {\bibfnamefont {T.}~\bibnamefont
  {Pluecker}}, \bibinfo {author} {\bibfnamefont {M.~R.}\ \bibnamefont
  {Wegewijs}}, \ and\ \bibinfo {author} {\bibfnamefont {J.}~\bibnamefont
  {Splettstoesser}},\ }\bibfield  {title} {\enquote {\bibinfo {title} {{Gauge
  freedom in observables and Landsberg's nonadiabatic geometric phase: Pumping
  spectroscopy of interacting open quantum systems}},}\ }\href {\doibase
  10.1103/PhysRevB.95.155431} {\bibfield  {journal} {\bibinfo  {journal} {Phys.
  Rev. B}\ }\textbf {\bibinfo {volume} {95}},\ \bibinfo {pages} {155431}
  (\bibinfo {year} {2017}{\natexlab{a}})}\BibitemShut {NoStop}%
\bibitem [{\citenamefont {Pluecker}\ \emph
  {et~al.}(2017{\natexlab{b}})\citenamefont {Pluecker}, \citenamefont
  {Wegewijs},\ and\ \citenamefont {Splettstoesser}}]{Pluecker2017Nov}%
  \BibitemOpen
  \bibfield  {author} {\bibinfo {author} {\bibfnamefont {T.}~\bibnamefont
  {Pluecker}}, \bibinfo {author} {\bibfnamefont {M.~R.}\ \bibnamefont
  {Wegewijs}}, \ and\ \bibinfo {author} {\bibfnamefont {J.}~\bibnamefont
  {Splettstoesser}},\ }\bibfield  {title} {\enquote {\bibinfo {title} {{Meter
  calibration and the geometric pumping process in open quantum systems}},}\
  }\href {https://arxiv.org/abs/1711.10431} {\bibfield  {journal} {\bibinfo
  {journal} {arXiv}\ } (\bibinfo {year} {2017}{\natexlab{b}})},\ \Eprint
  {http://arxiv.org/abs/1711.10431} {1711.10431} \BibitemShut {NoStop}%
\bibitem [{\citenamefont {Nakajima}\ \emph {et~al.}(2015)\citenamefont
  {Nakajima}, \citenamefont {Taguchi}, \citenamefont {Kubo},\ and\
  \citenamefont {Tokura}}]{Nakajima2015Nov}%
  \BibitemOpen
  \bibfield  {author} {\bibinfo {author} {\bibfnamefont {S.}~\bibnamefont
  {Nakajima}}, \bibinfo {author} {\bibfnamefont {M.}~\bibnamefont {Taguchi}},
  \bibinfo {author} {\bibfnamefont {T.}~\bibnamefont {Kubo}}, \ and\ \bibinfo
  {author} {\bibfnamefont {Y.}~\bibnamefont {Tokura}},\ }\bibfield  {title}
  {\enquote {\bibinfo {title} {{Interaction effect on adiabatic pump of charge
  and spin in quantum dot}},}\ }\href {\doibase 10.1103/PhysRevB.92.195420}
  {\bibfield  {journal} {\bibinfo  {journal} {Phys. Rev. B}\ }\textbf {\bibinfo
  {volume} {92}},\ \bibinfo {pages} {195420} (\bibinfo {year}
  {2015})}\BibitemShut {NoStop}%
\bibitem [{\citenamefont {Pekola}\ \emph {et~al.}(2013)\citenamefont {Pekola},
  \citenamefont {Saira}, \citenamefont {Maisi}, \citenamefont {Kemppinen},
  \citenamefont
  {M{\ifmmode\ddot{o}\else\"{o}\fi}tt{\ifmmode\ddot{o}\else\"{o}\fi}nen},
  \citenamefont {Pashkin},\ and\ \citenamefont {Averin}}]{Pekola2013Oct}%
  \BibitemOpen
  \bibfield  {author} {\bibinfo {author} {\bibfnamefont {J.~P.}\ \bibnamefont
  {Pekola}}, \bibinfo {author} {\bibfnamefont {O.-P.}\ \bibnamefont {Saira}},
  \bibinfo {author} {\bibfnamefont {V.~F.}\ \bibnamefont {Maisi}}, \bibinfo
  {author} {\bibfnamefont {A.}~\bibnamefont {Kemppinen}}, \bibinfo {author}
  {\bibfnamefont {M.}~\bibnamefont
  {M{\ifmmode\ddot{o}\else\"{o}\fi}tt{\ifmmode\ddot{o}\else\"{o}\fi}nen}},
  \bibinfo {author} {\bibfnamefont {Y.~A.}\ \bibnamefont {Pashkin}}, \ and\
  \bibinfo {author} {\bibfnamefont {D.~V.}\ \bibnamefont {Averin}},\ }\bibfield
   {title} {\enquote {\bibinfo {title} {{Single-electron current sources:
  Toward a refined definition of the ampere}},}\ }\href {\doibase
  10.1103/RevModPhys.85.1421} {\bibfield  {journal} {\bibinfo  {journal} {Rev.
  Mod. Phys.}\ }\textbf {\bibinfo {volume} {85}},\ \bibinfo {pages}
  {1421--1472} (\bibinfo {year} {2013})}\BibitemShut {NoStop}%
\bibitem [{\citenamefont {Battista}\ \emph
  {et~al.}(2014{\natexlab{a}})\citenamefont {Battista}, \citenamefont {Haupt},\
  and\ \citenamefont {Splettstoesser}}]{Battista2014Aug}%
  \BibitemOpen
  \bibfield  {author} {\bibinfo {author} {\bibfnamefont {F.}~\bibnamefont
  {Battista}}, \bibinfo {author} {\bibfnamefont {F.}~\bibnamefont {Haupt}}, \
  and\ \bibinfo {author} {\bibfnamefont {J.}~\bibnamefont {Splettstoesser}},\
  }\bibfield  {title} {\enquote {\bibinfo {title} {Energy and power
  fluctuations in ac-driven coherent conductors},}\ }\href {\doibase
  10.1103/PhysRevB.90.085418} {\bibfield  {journal} {\bibinfo  {journal} {Phys.
  Rev. B}\ }\textbf {\bibinfo {volume} {90}},\ \bibinfo {pages} {085418}
  (\bibinfo {year} {2014}{\natexlab{a}})}\BibitemShut {NoStop}%
\bibitem [{\citenamefont {Battista}\ \emph
  {et~al.}(2014{\natexlab{b}})\citenamefont {Battista}, \citenamefont {Haupt},\
  and\ \citenamefont {Splettstoesser}}]{Battista2014Dec}%
  \BibitemOpen
  \bibfield  {author} {\bibinfo {author} {\bibfnamefont {F.}~\bibnamefont
  {Battista}}, \bibinfo {author} {\bibfnamefont {F.}~\bibnamefont {Haupt}}, \
  and\ \bibinfo {author} {\bibfnamefont {J.}~\bibnamefont {Splettstoesser}},\
  }\bibfield  {title} {\enquote {\bibinfo {title} {{Correlations between charge
  and energy current in ac-driven coherent conductors}},}\ }\href {\doibase
  10.1088/1742-6596/568/5/052008} {\bibfield  {journal} {\bibinfo  {journal}
  {J. Phys. Conf. Ser.}\ }\textbf {\bibinfo {volume} {568}},\ \bibinfo {pages}
  {052008} (\bibinfo {year} {2014}{\natexlab{b}})}\BibitemShut {NoStop}%
\bibitem [{\citenamefont {Moskalets}(2014{\natexlab{a}})}]{Moskalets2014May}%
  \BibitemOpen
  \bibfield  {author} {\bibinfo {author} {\bibfnamefont {M.}~\bibnamefont
  {Moskalets}},\ }\bibfield  {title} {\enquote {\bibinfo {title} {{Floquet
  Scattering Matrix Theory of Heat Fluctuations in Dynamical Quantum
  Conductors}},}\ }\href {\doibase 10.1103/PhysRevLett.112.206801} {\bibfield
  {journal} {\bibinfo  {journal} {Phys. Rev. Lett.}\ }\textbf {\bibinfo
  {volume} {112}},\ \bibinfo {pages} {206801} (\bibinfo {year}
  {2014}{\natexlab{a}})}\BibitemShut {NoStop}%
\bibitem [{\citenamefont {Moskalets}(2014{\natexlab{b}})}]{Moskalets14Err}%
  \BibitemOpen
  \bibfield  {author} {\bibinfo {author} {\bibfnamefont {M.}~\bibnamefont
  {Moskalets}},\ }\bibfield  {title} {\enquote {\bibinfo {title} {Erratum:
  Floquet scattering matrix theory of heat fluctuations in dynamical quantum
  conductors [phys. rev. lett. \textbf{112} , 206801 (2014)]},}\ }\href
  {\doibase 10.1103/PhysRevLett.113.069902} {\bibfield  {journal} {\bibinfo
  {journal} {Phys. Rev. Lett.}\ }\textbf {\bibinfo {volume} {113}},\ \bibinfo
  {pages} {069902} (\bibinfo {year} {2014}{\natexlab{b}})}\BibitemShut
  {NoStop}%
\bibitem [{\citenamefont {Terr{\ifmmode\acute{e}\else\'{e}\fi}n~Alonso}\ \emph
  {et~al.}(2019)\citenamefont {Terr{\ifmmode\acute{e}\else\'{e}\fi}n~Alonso},
  \citenamefont {Romero},\ and\ \citenamefont
  {Arrachea}}]{TerrenAlonso2019Mar}%
  \BibitemOpen
  \bibfield  {author} {\bibinfo {author} {\bibfnamefont {P.}~\bibnamefont
  {Terr{\ifmmode\acute{e}\else\'{e}\fi}n~Alonso}}, \bibinfo {author}
  {\bibfnamefont {J.}~\bibnamefont {Romero}}, \ and\ \bibinfo {author}
  {\bibfnamefont {L.}~\bibnamefont {Arrachea}},\ }\bibfield  {title} {\enquote
  {\bibinfo {title} {{Work exchange, geometric magnetization, and
  fluctuation-dissipation relations in a quantum dot under adiabatic
  magnetoelectric driving}},}\ }\href {\doibase 10.1103/PhysRevB.99.115424}
  {\bibfield  {journal} {\bibinfo  {journal} {Phys. Rev. B}\ }\textbf {\bibinfo
  {volume} {99}},\ \bibinfo {pages} {115424} (\bibinfo {year}
  {2019})}\BibitemShut {NoStop}%
\bibitem [{\citenamefont {Popovic}\ \emph {et~al.}(2020)\citenamefont
  {Popovic}, \citenamefont {Mitchison}, \citenamefont {Strathearn},
  \citenamefont {Lovett}, \citenamefont {Goold},\ and\ \citenamefont
  {Eastham}}]{Popovic2020Aug}%
  \BibitemOpen
  \bibfield  {author} {\bibinfo {author} {\bibfnamefont {M.}~\bibnamefont
  {Popovic}}, \bibinfo {author} {\bibfnamefont {M.~T.}\ \bibnamefont
  {Mitchison}}, \bibinfo {author} {\bibfnamefont {A.}~\bibnamefont
  {Strathearn}}, \bibinfo {author} {\bibfnamefont {B.~W.}\ \bibnamefont
  {Lovett}}, \bibinfo {author} {\bibfnamefont {J.}~\bibnamefont {Goold}}, \
  and\ \bibinfo {author} {\bibfnamefont {P.~R.}\ \bibnamefont {Eastham}},\
  }\bibfield  {title} {\enquote {\bibinfo {title} {{Non-equilibrium quantum
  thermodynamics with time-evolving matrix product operators}},}\ }\href
  {https://arxiv.org/abs/2008.06491v1} {\bibfield  {journal} {\bibinfo
  {journal} {arXiv}\ } (\bibinfo {year} {2020})},\ \Eprint
  {http://arxiv.org/abs/2008.06491} {2008.06491} \BibitemShut {NoStop}%
\bibitem [{\citenamefont {Bagrets}\ and\ \citenamefont
  {Nazarov}(2003)}]{Bagrets2003Feb}%
  \BibitemOpen
  \bibfield  {author} {\bibinfo {author} {\bibfnamefont {D.~A.}\ \bibnamefont
  {Bagrets}}\ and\ \bibinfo {author} {\bibfnamefont {{\relax Yu}.~V.}\
  \bibnamefont {Nazarov}},\ }\bibfield  {title} {\enquote {\bibinfo {title}
  {{Full counting statistics of charge transfer in Coulomb blockade
  systems}},}\ }\href {\doibase 10.1103/PhysRevB.67.085316} {\bibfield
  {journal} {\bibinfo  {journal} {Phys. Rev. B}\ }\textbf {\bibinfo {volume}
  {67}},\ \bibinfo {pages} {085316} (\bibinfo {year} {2003})}\BibitemShut
  {NoStop}%
\bibitem [{\citenamefont {Mostafazadeh}(1997)}]{Mostafazadeh1997Mar}%
  \BibitemOpen
  \bibfield  {author} {\bibinfo {author} {\bibfnamefont {A.}~\bibnamefont
  {Mostafazadeh}},\ }\bibfield  {title} {\enquote {\bibinfo {title} {Quantum
  adiabatic approximation and the geometric phase},}\ }\href {\doibase
  10.1103/PhysRevA.55.1653} {\bibfield  {journal} {\bibinfo  {journal} {Phys.
  Rev. A}\ }\textbf {\bibinfo {volume} {55}},\ \bibinfo {pages} {1653--1664}
  (\bibinfo {year} {1997})}\BibitemShut {NoStop}%
\bibitem [{\citenamefont {Wunsch}\ \emph {et~al.}(2005)\citenamefont {Wunsch},
  \citenamefont {Braun}, \citenamefont {K{\ifmmode\ddot{o}\else\"{o}\fi}nig},\
  and\ \citenamefont {Pfannkuche}}]{Wunsch2005Nov}%
  \BibitemOpen
  \bibfield  {author} {\bibinfo {author} {\bibfnamefont {B.}~\bibnamefont
  {Wunsch}}, \bibinfo {author} {\bibfnamefont {M.}~\bibnamefont {Braun}},
  \bibinfo {author} {\bibfnamefont {J.}~\bibnamefont
  {K{\ifmmode\ddot{o}\else\"{o}\fi}nig}}, \ and\ \bibinfo {author}
  {\bibfnamefont {D.}~\bibnamefont {Pfannkuche}},\ }\bibfield  {title}
  {\enquote {\bibinfo {title} {{Probing level renormalization by sequential
  transport through double quantum dots}},}\ }\href {\doibase
  10.1103/PhysRevB.72.205319} {\bibfield  {journal} {\bibinfo  {journal} {Phys.
  Rev. B}\ }\textbf {\bibinfo {volume} {72}},\ \bibinfo {pages} {205319}
  (\bibinfo {year} {2005})}\BibitemShut {NoStop}%
\bibitem [{\citenamefont {Braun}\ \emph {et~al.}(2004)\citenamefont {Braun},
  \citenamefont {K{\ifmmode\ddot{o}\else\"{o}\fi}nig},\ and\ \citenamefont
  {Martinek}}]{Braun2004Nov}%
  \BibitemOpen
  \bibfield  {author} {\bibinfo {author} {\bibfnamefont {M.}~\bibnamefont
  {Braun}}, \bibinfo {author} {\bibfnamefont {J.}~\bibnamefont
  {K{\ifmmode\ddot{o}\else\"{o}\fi}nig}}, \ and\ \bibinfo {author}
  {\bibfnamefont {J.}~\bibnamefont {Martinek}},\ }\bibfield  {title} {\enquote
  {\bibinfo {title} {{Theory of transport through quantum-dot spin valves in
  the weak-coupling regime}},}\ }\href {\doibase 10.1103/PhysRevB.70.195345}
  {\bibfield  {journal} {\bibinfo  {journal} {Phys. Rev. B}\ }\textbf {\bibinfo
  {volume} {70}},\ \bibinfo {pages} {195345} (\bibinfo {year}
  {2004})}\BibitemShut {NoStop}%
\bibitem [{\citenamefont {Cavaliere}\ \emph {et~al.}(2009)\citenamefont
  {Cavaliere}, \citenamefont {Governale},\ and\ \citenamefont
  {K{\ifmmode\ddot{o}\else\"{o}\fi}nig}}]{Cavaliere2009Sep}%
  \BibitemOpen
  \bibfield  {author} {\bibinfo {author} {\bibfnamefont {F.}~\bibnamefont
  {Cavaliere}}, \bibinfo {author} {\bibfnamefont {M.}~\bibnamefont
  {Governale}}, \ and\ \bibinfo {author} {\bibfnamefont {J.}~\bibnamefont
  {K{\ifmmode\ddot{o}\else\"{o}\fi}nig}},\ }\bibfield  {title} {\enquote
  {\bibinfo {title} {{Nonadiabatic Pumping through Interacting Quantum
  Dots}},}\ }\href {\doibase 10.1103/PhysRevLett.103.136801} {\bibfield
  {journal} {\bibinfo  {journal} {Phys. Rev. Lett.}\ }\textbf {\bibinfo
  {volume} {103}},\ \bibinfo {pages} {136801} (\bibinfo {year}
  {2009})}\BibitemShut {NoStop}%
\bibitem [{\citenamefont {Saptsov}\ and\ \citenamefont
  {Wegewijs}(2012)}]{Saptsov2012Dec}%
  \BibitemOpen
  \bibfield  {author} {\bibinfo {author} {\bibfnamefont {R.~B.}\ \bibnamefont
  {Saptsov}}\ and\ \bibinfo {author} {\bibfnamefont {M.~R.}\ \bibnamefont
  {Wegewijs}},\ }\bibfield  {title} {\enquote {\bibinfo {title} {{Fermionic
  superoperators for zero-temperature nonlinear transport: Real-time
  perturbation theory and renormalization group for Anderson quantum dots}},}\
  }\href {\doibase 10.1103/PhysRevB.86.235432} {\bibfield  {journal} {\bibinfo
  {journal} {Phys. Rev. B}\ }\textbf {\bibinfo {volume} {86}},\ \bibinfo
  {pages} {235432} (\bibinfo {year} {2012})}\BibitemShut {NoStop}%
\bibitem [{\citenamefont {Saptsov}\ and\ \citenamefont
  {Wegewijs}(2014)}]{Saptsov2014Jul}%
  \BibitemOpen
  \bibfield  {author} {\bibinfo {author} {\bibfnamefont {R.~B.}\ \bibnamefont
  {Saptsov}}\ and\ \bibinfo {author} {\bibfnamefont {M.~R.}\ \bibnamefont
  {Wegewijs}},\ }\bibfield  {title} {\enquote {\bibinfo {title}
  {{Time-dependent quantum transport: Causal superfermions, exact
  fermion-parity protected decay modes, and Pauli exclusion principle for mixed
  quantum states}},}\ }\href {\doibase 10.1103/PhysRevB.90.045407} {\bibfield
  {journal} {\bibinfo  {journal} {Phys. Rev. B}\ }\textbf {\bibinfo {volume}
  {90}},\ \bibinfo {pages} {045407} (\bibinfo {year} {2014})}\BibitemShut
  {NoStop}%
\bibitem [{\citenamefont {Schulenborg}\ \emph {et~al.}(2014)\citenamefont
  {Schulenborg}, \citenamefont {Splettstoesser}, \citenamefont {Governale},\
  and\ \citenamefont {Contreras-Pulido}}]{Schulenborg2014May}%
  \BibitemOpen
  \bibfield  {author} {\bibinfo {author} {\bibfnamefont {J.}~\bibnamefont
  {Schulenborg}}, \bibinfo {author} {\bibfnamefont {J.}~\bibnamefont
  {Splettstoesser}}, \bibinfo {author} {\bibfnamefont {M.}~\bibnamefont
  {Governale}}, \ and\ \bibinfo {author} {\bibfnamefont {L.~D.}\ \bibnamefont
  {Contreras-Pulido}},\ }\bibfield  {title} {\enquote {\bibinfo {title}
  {{Detection of the relaxation rates of an interacting quantum dot by a
  capacitively coupled sensor dot}},}\ }\href {\doibase
  10.1103/PhysRevB.89.195305} {\bibfield  {journal} {\bibinfo  {journal} {Phys.
  Rev. B}\ }\textbf {\bibinfo {volume} {89}},\ \bibinfo {pages} {195305}
  (\bibinfo {year} {2014})}\BibitemShut {NoStop}%
\bibitem [{\citenamefont {Splettstoesser}\ \emph {et~al.}(2006)\citenamefont
  {Splettstoesser}, \citenamefont {Governale}, \citenamefont
  {K{\ifmmode\ddot{o}\else\"{o}\fi}nig},\ and\ \citenamefont
  {Fazio}}]{Splettstoesser2006Aug}%
  \BibitemOpen
  \bibfield  {author} {\bibinfo {author} {\bibfnamefont {J.}~\bibnamefont
  {Splettstoesser}}, \bibinfo {author} {\bibfnamefont {M.}~\bibnamefont
  {Governale}}, \bibinfo {author} {\bibfnamefont {J.}~\bibnamefont
  {K{\ifmmode\ddot{o}\else\"{o}\fi}nig}}, \ and\ \bibinfo {author}
  {\bibfnamefont {R.}~\bibnamefont {Fazio}},\ }\bibfield  {title} {\enquote
  {\bibinfo {title} {{Adiabatic pumping through a quantum dot with coulomb
  interactions: A perturbation expansion in the tunnel coupling}},}\ }\href
  {\doibase 10.1103/PhysRevB.74.085305} {\bibfield  {journal} {\bibinfo
  {journal} {Phys. Rev. B}\ }\textbf {\bibinfo {volume} {74}},\ \bibinfo
  {pages} {085305} (\bibinfo {year} {2006})}\BibitemShut {NoStop}%
\bibitem [{\citenamefont {Reckermann}\ \emph {et~al.}(2010)\citenamefont
  {Reckermann}, \citenamefont {Splettstoesser},\ and\ \citenamefont
  {Wegewijs}}]{Reckermann2010Jun}%
  \BibitemOpen
  \bibfield  {author} {\bibinfo {author} {\bibfnamefont {F.}~\bibnamefont
  {Reckermann}}, \bibinfo {author} {\bibfnamefont {J.}~\bibnamefont
  {Splettstoesser}}, \ and\ \bibinfo {author} {\bibfnamefont {M.~R.}\
  \bibnamefont {Wegewijs}},\ }\bibfield  {title} {\enquote {\bibinfo {title}
  {Interaction-induced adiabatic nonlinear transport},}\ }\href {\doibase
  10.1103/PhysRevLett.104.226803} {\bibfield  {journal} {\bibinfo  {journal}
  {Phys. Rev. Lett.}\ }\textbf {\bibinfo {volume} {104}},\ \bibinfo {pages}
  {226803} (\bibinfo {year} {2010})}\BibitemShut {NoStop}%
\bibitem [{\citenamefont {B{\ifmmode\ddot{u}\else\"{u}\fi}ttiker}\ \emph
  {et~al.}(1994)\citenamefont {B{\ifmmode\ddot{u}\else\"{u}\fi}ttiker},
  \citenamefont {Thomas},\ and\ \citenamefont
  {Pr{\ifmmode\hat{e}\else\^{e}\fi}tre}}]{Buttiker1994Mar}%
  \BibitemOpen
  \bibfield  {author} {\bibinfo {author} {\bibfnamefont {M.}~\bibnamefont
  {B{\ifmmode\ddot{u}\else\"{u}\fi}ttiker}}, \bibinfo {author} {\bibfnamefont
  {H.}~\bibnamefont {Thomas}}, \ and\ \bibinfo {author} {\bibfnamefont
  {A.}~\bibnamefont {Pr{\ifmmode\hat{e}\else\^{e}\fi}tre}},\ }\bibfield
  {title} {\enquote {\bibinfo {title} {{Current partition in multiprobe
  conductors in the presence of slowly oscillating external potentials}},}\
  }\href {\doibase 10.1007/BF01307664} {\bibfield  {journal} {\bibinfo
  {journal} {Z. Phys. B: Condens. Matter}\ }\textbf {\bibinfo {volume} {94}},\
  \bibinfo {pages} {133--137} (\bibinfo {year} {1994})}\BibitemShut {NoStop}%
\bibitem [{\citenamefont {Brouwer}(2001)}]{Brouwer2001Mar}%
  \BibitemOpen
  \bibfield  {author} {\bibinfo {author} {\bibfnamefont {P.~W.}\ \bibnamefont
  {Brouwer}},\ }\bibfield  {title} {\enquote {\bibinfo {title} {Rectification
  of displacement currents in an adiabatic electron pump},}\ }\href {\doibase
  10.1103/PhysRevB.63.121303} {\bibfield  {journal} {\bibinfo  {journal} {Phys.
  Rev. B}\ }\textbf {\bibinfo {volume} {63}},\ \bibinfo {pages} {121303}
  (\bibinfo {year} {2001})}\BibitemShut {NoStop}%
\bibitem [{\citenamefont {Avron}\ \emph {et~al.}(2004)\citenamefont {Avron},
  \citenamefont {Elgart}, \citenamefont {Graf},\ and\ \citenamefont
  {Sadun}}]{Avron2004Aug}%
  \BibitemOpen
  \bibfield  {author} {\bibinfo {author} {\bibfnamefont {J.~E.}\ \bibnamefont
  {Avron}}, \bibinfo {author} {\bibfnamefont {A.}~\bibnamefont {Elgart}},
  \bibinfo {author} {\bibfnamefont {G.~M.}\ \bibnamefont {Graf}}, \ and\
  \bibinfo {author} {\bibfnamefont {L.}~\bibnamefont {Sadun}},\ }\bibfield
  {title} {\enquote {\bibinfo {title} {{Transport and Dissipation in Quantum
  Pumps}},}\ }\href {\doibase 10.1023/B:JOSS.0000037245.45780.e1} {\bibfield
  {journal} {\bibinfo  {journal} {J. Stat. Phys.}\ }\textbf {\bibinfo {volume}
  {116}},\ \bibinfo {pages} {425--473} (\bibinfo {year} {2004})}\BibitemShut
  {NoStop}%
\bibitem [{\citenamefont {Splettstoesser}\ \emph {et~al.}(2005)\citenamefont
  {Splettstoesser}, \citenamefont {Governale}, \citenamefont
  {K{\ifmmode\ddot{o}\else\"{o}\fi}nig},\ and\ \citenamefont
  {Fazio}}]{Splettstoesser2005Dec}%
  \BibitemOpen
  \bibfield  {author} {\bibinfo {author} {\bibfnamefont {J.}~\bibnamefont
  {Splettstoesser}}, \bibinfo {author} {\bibfnamefont {M.}~\bibnamefont
  {Governale}}, \bibinfo {author} {\bibfnamefont {J.}~\bibnamefont
  {K{\ifmmode\ddot{o}\else\"{o}\fi}nig}}, \ and\ \bibinfo {author}
  {\bibfnamefont {R.}~\bibnamefont {Fazio}},\ }\bibfield  {title} {\enquote
  {\bibinfo {title} {{Adiabatic Pumping through Interacting Quantum Dots}},}\
  }\href {\doibase 10.1103/PhysRevLett.95.246803} {\bibfield  {journal}
  {\bibinfo  {journal} {Phys. Rev. Lett.}\ }\textbf {\bibinfo {volume} {95}},\
  \bibinfo {pages} {246803} (\bibinfo {year} {2005})}\BibitemShut {NoStop}%
\bibitem [{\citenamefont {Moskalets}\ and\ \citenamefont
  {B{\ifmmode\ddot{u}\else\"{u}\fi}ttiker}(2002)}]{Moskalets2002Nov}%
  \BibitemOpen
  \bibfield  {author} {\bibinfo {author} {\bibfnamefont {M.}~\bibnamefont
  {Moskalets}}\ and\ \bibinfo {author} {\bibfnamefont {M.}~\bibnamefont
  {B{\ifmmode\ddot{u}\else\"{u}\fi}ttiker}},\ }\bibfield  {title} {\enquote
  {\bibinfo {title} {{Floquet scattering theory of quantum pumps}},}\ }\href
  {\doibase 10.1103/PhysRevB.66.205320} {\bibfield  {journal} {\bibinfo
  {journal} {Phys. Rev. B}\ }\textbf {\bibinfo {volume} {66}},\ \bibinfo
  {pages} {205320} (\bibinfo {year} {2002})}\BibitemShut {NoStop}%
\bibitem [{\citenamefont {Gasparinetti}\ \emph {et~al.}(2015)\citenamefont
  {Gasparinetti}, \citenamefont {Viisanen}, \citenamefont {Saira},
  \citenamefont {Faivre}, \citenamefont {Arzeo}, \citenamefont {Meschke},\ and\
  \citenamefont {Pekola}}]{Gasparinetti2015Jan}%
  \BibitemOpen
  \bibfield  {author} {\bibinfo {author} {\bibfnamefont {S.}~\bibnamefont
  {Gasparinetti}}, \bibinfo {author} {\bibfnamefont {K.~L.}\ \bibnamefont
  {Viisanen}}, \bibinfo {author} {\bibfnamefont {O.-P.}\ \bibnamefont {Saira}},
  \bibinfo {author} {\bibfnamefont {T.}~\bibnamefont {Faivre}}, \bibinfo
  {author} {\bibfnamefont {M.}~\bibnamefont {Arzeo}}, \bibinfo {author}
  {\bibfnamefont {M.}~\bibnamefont {Meschke}}, \ and\ \bibinfo {author}
  {\bibfnamefont {J.~P.}\ \bibnamefont {Pekola}},\ }\bibfield  {title}
  {\enquote {\bibinfo {title} {{Fast Electron Thermometry for Ultrasensitive
  Calorimetric Detection}},}\ }\href {\doibase 10.1103/PhysRevApplied.3.014007}
  {\bibfield  {journal} {\bibinfo  {journal} {Phys. Rev. Appl.}\ }\textbf
  {\bibinfo {volume} {3}},\ \bibinfo {pages} {014007} (\bibinfo {year}
  {2015})}\BibitemShut {NoStop}%
\bibitem [{\citenamefont {Wang}\ \emph {et~al.}(2018)\citenamefont {Wang},
  \citenamefont {Saira},\ and\ \citenamefont {Pekola}}]{Wang2018Jan}%
  \BibitemOpen
  \bibfield  {author} {\bibinfo {author} {\bibfnamefont {L.~B.}\ \bibnamefont
  {Wang}}, \bibinfo {author} {\bibfnamefont {O.-P.}\ \bibnamefont {Saira}}, \
  and\ \bibinfo {author} {\bibfnamefont {J.~P.}\ \bibnamefont {Pekola}},\
  }\bibfield  {title} {\enquote {\bibinfo {title} {{Fast thermometry with a
  proximity Josephson junction}},}\ }\href {\doibase 10.1063/1.5010236}
  {\bibfield  {journal} {\bibinfo  {journal} {Appl. Phys. Lett.}\ }\textbf
  {\bibinfo {volume} {112}},\ \bibinfo {pages} {013105} (\bibinfo {year}
  {2018})}\BibitemShut {NoStop}%
\bibitem [{\citenamefont {Placke}\ \emph {et~al.}(2018)\citenamefont {Placke},
  \citenamefont {Pluecker}, \citenamefont {Splettstoesser},\ and\ \citenamefont
  {Wegewijs}}]{Placke2018Aug}%
  \BibitemOpen
  \bibfield  {author} {\bibinfo {author} {\bibfnamefont {B.~A.}\ \bibnamefont
  {Placke}}, \bibinfo {author} {\bibfnamefont {T.}~\bibnamefont {Pluecker}},
  \bibinfo {author} {\bibfnamefont {J.}~\bibnamefont {Splettstoesser}}, \ and\
  \bibinfo {author} {\bibfnamefont {M.~R.}\ \bibnamefont {Wegewijs}},\
  }\bibfield  {title} {\enquote {\bibinfo {title} {{Attractive and driven
  interactions in quantum dots: Mechanisms for geometric pumping}},}\ }\href
  {\doibase 10.1103/PhysRevB.98.085307} {\bibfield  {journal} {\bibinfo
  {journal} {Phys. Rev. B}\ }\textbf {\bibinfo {volume} {98}},\ \bibinfo
  {pages} {085307} (\bibinfo {year} {2018})}\BibitemShut {NoStop}%
\bibitem [{\citenamefont {Hamo}\ \emph {et~al.}(2016)\citenamefont {Hamo},
  \citenamefont {Benyamini}, \citenamefont {Shapir}, \citenamefont {Khivrich},
  \citenamefont {Waissman}, \citenamefont {Kaasbjerg}, \citenamefont {Oreg},
  \citenamefont {von Oppen},\ and\ \citenamefont {Ilani}}]{Hamo2016Jul}%
  \BibitemOpen
  \bibfield  {author} {\bibinfo {author} {\bibfnamefont {A.}~\bibnamefont
  {Hamo}}, \bibinfo {author} {\bibfnamefont {A.}~\bibnamefont {Benyamini}},
  \bibinfo {author} {\bibfnamefont {I.}~\bibnamefont {Shapir}}, \bibinfo
  {author} {\bibfnamefont {I.}~\bibnamefont {Khivrich}}, \bibinfo {author}
  {\bibfnamefont {J.}~\bibnamefont {Waissman}}, \bibinfo {author}
  {\bibfnamefont {K.}~\bibnamefont {Kaasbjerg}}, \bibinfo {author}
  {\bibfnamefont {Y.}~\bibnamefont {Oreg}}, \bibinfo {author} {\bibfnamefont
  {F.}~\bibnamefont {von Oppen}}, \ and\ \bibinfo {author} {\bibfnamefont
  {S.}~\bibnamefont {Ilani}},\ }\bibfield  {title} {\enquote {\bibinfo {title}
  {{Electron attraction mediated by Coulomb repulsion}},}\ }\href {\doibase
  10.1038/nature18639} {\bibfield  {journal} {\bibinfo  {journal} {Nature}\
  }\textbf {\bibinfo {volume} {535}},\ \bibinfo {pages} {395--400} (\bibinfo
  {year} {2016})}\BibitemShut {NoStop}%
\end{thebibliography}%

\end{document}